\documentclass[usenatbib,onecolumn]{mn2e}
\usepackage{graphicx}
\usepackage{amssymb}
\usepackage{amsmath}
\usepackage{times}

\title[Electromagnetic Fields in the Exterior of an Oscillating
Relativistic Star I.]
{Electromagnetic Fields in the Exterior of an Oscillating Relativistic
Star -- I.  General Expressions and application to a rotating magnetic
dipole}

\author[L.~Rezzolla, B.~J.~Ahmedov]
        {Luciano~Rezzolla$^{(1,\;2)}$, Bobomurat~J.~Ahmedov$^{(3,\;4)}$
               						        \\
                                                                \\
        $^{(1)}$SISSA, International School for Advanced Studies, and
        INFN--Trieste, Via Beirut 2-4, 34014 Trieste, Italy     \\
        $^{(2)}$Department of Physics and Astronomy, Louisiana State
        University, Baton Rouge, LA 70803 USA \\
        $^{(3)}$Ulugh Beg Astronomical Institute,
        Astronomicheskaya 33, Tashkent 700052, Uzbekistan       \\
        $^{(4)}$Institute of Nuclear Physics,
        Ulughbek, Tashkent 702132, Uzbekistan                   \\
                \\}
\pagerange{\pageref{firstpage}--\pageref{lastpage}}

\pubyear{2004}

\begin{document}

\maketitle

\label{firstpage}

\begin{abstract}
	Relativistic  stars  are  endowed  with  intense  electromagnetic
	fields but are also subject to oscillations of various types.  We
	here  investigate  the  impact  that  oscillations  have  on  the
	electric and  magnetic fields external to a  relativistic star in
	vacuum.   In particular,  modelling  the star  as a  relativistic
	polytrope with infinite conductivity, we consider the solution of
	the general  relativistic Maxwell equations both  in the vicinity
	of  the stellar  surface and  far  from it,  once a  perturbative
	velocity  field is  specified.  In  this first  paper  we present
	general  analytic  expressions  that  are not  specialized  to  a
	particular magnetic  field topology or  velocity field.  However,
	as  a   validating  example  and   an  astrophysically  important
	application,  we  consider  a  dipolar  magnetic  field  and  the
	velocity field  corresponding to  the rotation of  the misaligned
	dipole.    Besides  providing   analytic   expressions  for   the
	electromagnetic  fields   produced  by  this   configuration,  we
	calculate, for  the first  time, the general  relativistic energy
	loss through dipolar electromagnetic  radiation. We find that the
	widely used  Newtonian expression under-estimates this  loss by a
	factor  of  2--6  depending  on the  stellar  compactness.   This
	correction could have important  consequences in the study of the
	spin evolution of pulsars.
\end{abstract}


\date{Accepted 0000 00 00.
      Received 0000 00 00.}

\section{Introduction}

	The study of the electrodynamics of relativistic stars when these
are undergoing normal-mode oscillations has the prospect of providing
several pieces of information which are relevant not only for the
astrophysics but also for the physics of these objects. Astronomical
observations indicate, in fact, that compact relativistic stars are
endowed with extremely intense magnetic fields which reach surface
strengths of the order of $\lesssim 10^{10}$ G in older neutron stars
associated with recycled pulsars and low-mass X-ray binaries.
Observations of young neutron stars, on the other hand, show surface
magnetic fields that are much stronger and usually of the order of
\hbox{$10^{11}-10^{13}$ G}. In addition to this, the phenomenology
associated with soft gamma repeaters and anomalous X-ray pulsars suggests
that the surface magnetic fields can become even stronger and up to
$10^{14}-10^{15}$ G (Mereghetti \& Stella 1995, Kouveliotou et al. 1998,
1999), where they could be produced through efficient dynamo processes
\citep{td95,bru03}.

	However, intense electromagnetic fields are not the only
important feature of compact objects. The same astronomical observations
that indicate the presence of such fields, in fact, also make it manifest
that these are associated with objects that are very compact and hence
with strong gravitational fields. As a result, an accurate description of
the electrodynamics of compact objects can only be made within a
framework in which the relativistic corrections are properly accounted
for. The literature investigating electromagnetic ``test-fields'' in
curved spacetimes ({\it i.e.} fields whose energy-density is small when
compared with the average rest-mass energy density) is rather extensive
and has considered both spherically symmetric spacetimes \citep{g1, a1}
and also slowly rotating spacetimes \citep{m1,m2,k3}. More recently,
\citet{r1,r2} have studied the interior and vacuum exterior
electromagnetic fields produced by a rotating dipole moment comoving with
the star and started a systematic approach to provide analytic
expressions for the regions of the spacetime near the stellar surface
({\it i.e.} the so-called {\it ``near-zone''}). These results have also
been extended beyond the low-frequency approximation by \citet{kmo03},
who have resorted to a numerical approach to find solutions to the
Maxwell equations that are approximate, but valid also at large distances
({\it i.e.} the so-called {\it ``wave-zone''}).

	If it is natural to expect strong electromagnetic fields in the
vicinity of relativistic stars, it is equally natural to expect that
these stars too, just like ordinary stars, will be subject to a variety
of oscillation modes that become manifest through the emission of both
gravitational waves and electromagnetic waves. Because both of these
observational windows may soon provide data that could reach us
simultaneously, they offer unique opportunities to test the physics and
internal structure of these objects as well as the properties of matter
at nuclear densities. The investigation of the electromagnetic fields
produced by an oscillating magnetized star is, however, a rather complex
problem and this is testified by the scarce literature produced so far in
this context. Exceptions in this sense are, to the best of our knowledge,
the works of McDermott et al. (1984) and Muslimov \& Tsygan (1986). While
the first authors (McDermott et al. 1984) have modelled the electric
fields produced in the wave-zone by a neutron star subject to both
toroidal and spheroidal modes to calculate the electromagnetic damping of
these oscillations, Muslimov \& Tsygan (1986) have instead computed the
exact analytical solutions for the electromagnetic fields produced, both
in the near-zone and in the wave-zone, by an oscillating neutron star
with a dipolar surface magnetic field. In both cases the neutron star was
considered in vacuum and, for simplicity, the stellar magnetic field was
considered unperturbed. More recently, Timokhin et al. (2000), have
dropped the assumption of a vacuum region outside the stellar surface and
performed an extensive investigation of the electrodynamics of the
magnetosphere when this is perturbed as a result of toroidal
oscillations.

	An aspect that these three works share is the use of a Newtonian
description of gravity and, as we have argued above, this is expected to
be a poor approximation in the vicinity of the stellar surface.  The aim
of the present work is to layout a general formalism for the calculation
of electromagnetic ``test-fields'' in the curved spacetime of a spherical
relativistic star subject to perturbations. We do this by extending the
approach developed for the electrodynamics of rotating relativistic stars
(Rezzolla et al. 2001a, b) to include the possibility that the conducting
crust may have a velocity field as a result of the stellar
oscillations. As in our previous investigations \citep{r1,r2}, our main
aim here is to provide analytic expressions in a form which is as simple
as possible and can therefore be used also in astrophysical
applications. In this first paper we will concentrate on the development
of the mathematical formalism and in particular on the solution in the
two distinct regions of the space for which exact solutions can be found,
{\it i.e.} near the stellar surface where the relativistic corrections
are most important (near-zone), and at very large distances from it where
the spacetime approaches a flat one (wave-zone). In both cases we find
that two main corrections emerge in the expressions for the relativistic
electromagnetic fields and these are associated: {\it i)} to the
amplification of the electromagnetic fields on the stellar surface
produced by the local nonzero spacetime curvature; {\it ii)} to the
gravitational redshift which will affect the electromagnetic waves as
these propagate in the curved spacetime.

	The expressions for the components of the electromagnetic fields
presented here for the two regimes are completely general and do not
refer to a specific magnetic field topology nor to a precise velocity
field. The application of these solutions to the case of those spheroidal
and toroidal velocity fields most frequently encountered in the study of
stellar oscillations will be presented in a companion paper
\citep{r3}. Here, however, to validate the expressions derived and to
provide a comparison with known results, we consider a background
magnetic dipole and the simplest of the perturbation velocity fields: the
one produced by the uniform rotation of the dipole. In the near-zone, the
comparison of the derived expressions for the electric and magnetic
fields can be compared with the solutions calculated by \citet{r1,kmo03},
while, in the wave-zone, the new components of the electromagnetic fields
can be used to calculate the general relativistic expression for the
electromagnetic energy loss through dipolar radiation. This validating
test is also useful to reveal that the well-known Newtonian expression
commonly used to estimate the rotational energy loss through dipolar
radiation in pulsars \citep{p68,go69,st83} under-estimates this loss by a
factor 6 if the star is very compact or by a factor of 2 if the star is
less compact.

	The paper is organized as follows: in Section \ref{basic} we
formulate the basic Maxwell equations in the exterior Schwarzschild
spacetime, while in Section~\ref{em_si} we discuss the expressions for
the electromagnetic fields in the stellar interior and the boundary
conditions that constraints them at the stellar surface when this is
infinitely conducting and subject to velocity perturbations.  Sections
\ref{em_nz} and \ref{em_wz} are devoted to the derivation of the vacuum
electromagnetic fields in the vicinity of the source and at large
distances from it. There, we provide the exact analytical solutions to
the general relativistic Maxwell equations for arbitrary background
magnetic fields and velocity perturbations. Section~\ref{rmd} discusses
how the formalism developed in the previous Section can find a simple and
yet useful application in the case of a rotating magnetic dipole and
provides the general relativistic correction to the well-known expression
for the energy loss through dipolar electromagnetic
radiation. Section~\ref{conclusions} contains our conclusions and the
prospects for further applications of the formalism presented here. A
number of details about the calculations carried out in the main part of
the paper are finally presented in the Appendices A--D.

        We here use a spacelike signature $(-,+,+,+)$ and a system of
units in which $c = 1$ (unless explicitly shown otherwise for
convenience). Greek indices are taken to run from 0 to 3, Latin indices
from 1 to 3, and we adopt the standard convention for the summation over
repeated indices. Furthermore we will indicate four-vectors with bold
symbols ({\it e.g.} ${\boldsymbol u}$) and three-vectors with an arrow
({\it e.g.} ${\vec u}$).

\section{Maxwell Equations in a Schwarzschild Spacetime}
\label{basic}

    The investigation of the response of a relativistic {\it and}
magnetized star upon a generic perturbation would require, in principle,
the use of the coupled system of the Maxwell {\it and} Einstein
equations. Such a formulation, however, is overly complicated for most
astrophysical applications and a few reasonable assumptions can be made
to simplify the treatment to a more tractable form. The first of these
assumptions concerns the corrections to the Einstein field equations
introduced through the mass-energy contribution of the electromagnetic
fields inside and outside the relativistic star. It is not difficult to
show that these electromagnetic corrections are proportional to the
electromagnetic energy density and are rather small in most compact and
magnetized stars. Indeed, if $\langle\rho_0\rangle$ is the average
rest-mass density of a star of mass $M$ and radius $R$ as measured at
infinity, these corrections are at most
\begin{equation}
\frac{B^2}{8\pi \langle\rho_0\rangle c^2} \simeq
	2.2\times 10^{-7}\left(\frac{B}{10^{15}~{\rm G}}\right)^2
	\left(\frac{1.4~M_{\odot}}{M}\right)
	\left(\frac{R}{15~{\rm Km}}\right)^3 \ ,
\end{equation}
when compared with the total rest-mass energy density. Hereafter we will
assume these corrections to be negligible for the stellar objects which
we are interested in and thus treat the electromagnetic fields as
``test-fields'' in a given curved background. Other authors, interested
also in the structural modifications produced in very highly magnetized
stars (Bocquet et al. 1995, Konno et al. 1999, 2000, Oron 2002, Ioka \&
Sasaki 2004), have preferred not to make this approximation and have
instead included the magnetic contribution to the spacetime
curvature. The second simplifying assumption is that we will consider
also negligible the corrections produced by a global rotation of the
spacetime and induced by the rotation of the compact star (these
corrections were discussed by Rezzolla et al. 2001a,b). The numerical
analyses carried out by Geppert et al. (2000) and Zanotti \& Rezzolla
(2002) show that this is a rather good approximation in most cases of
astrophysical relevance.

	As a result of these two assumptions we can work in the
background spacetime of a nonrotating star, whose line element in a
spherical coordinate system $(t,r,\theta ,\phi)$ is given by
\begin{equation}
\label{schw}
ds^2 = g_{00}(r) dt^2 + g_{11}(r) dr^2+
        r^2 d\theta ^2+ r^2\sin^2\theta d\phi ^2 \ .
\end{equation}
The portion of the spacetime external to the star ({\it i.e.} for $r \ge
R$) is simply given by the Schwarzschild solution with $-g_{00} = N^2
\equiv (1 - 2M/r)$ and $g_{11} = 1/g_{00}$. For the portion of the
spacetime interior to the star, on the other hand, the metric functions
can be specified in terms of two potentials $\Lambda(r)$ and $\Phi(r)$ so
that
\begin{equation}
g_{00} = -e^{2 \Phi(r)} \ ,
\hskip 0.5 cm
g_{11} = e^{2 \Lambda(r)} = \left(1-\frac{2 m(r)}{r}\right)^{-1}
    \ ,
\end{equation}
where $\rho(r)$ is the total energy density and $m(r)\equiv 4\pi\int^R_0
r^2 \rho(r) dr $ its coordinate volume integral. The precise form of
these potentials is obtained through the familiar solution of the
Einstein equations for relativistic spherical stars, {\it i.e.} the TOV
equations (Shapiro \& Teukolsky 1983) and the metric functions are then
matched continuously to the external Schwarzschild spacetime so that
\begin{equation}
- g_{00}(r=R) = N_{_R}^2 \equiv 1 - \frac{2M}{R}\ ,
	\hskip 2.0 cm {\rm and}  \hskip 2.0 cm 
	g_{11}(r=R) = \frac{1}{N_{_R}^2} \ .
\end{equation}

    Within the external portion of this background spacetime we select a
family of static observers with four-velocity components
\begin{equation}
\label{obs}
(u^{\alpha})_{_{\rm obs}}\equiv
        N^{-1}\bigg(1,0,0,0\bigg) \ ,
        \hskip 2.0cm
(u_{\alpha})_{_{\rm obs}}\equiv
        N \bigg(- 1,0,0,0 \bigg) \ ,
\end{equation}
and associate to them orthonormal frames with tetrad four-vectors
${\boldsymbol e}_{\hat \mu} = ({\boldsymbol e}_{\hat 0}, {\boldsymbol
e}_{\hat r}, {\boldsymbol e}_{\hat \theta}, {\boldsymbol e}_{\hat \phi})$
and 1-forms ${\boldsymbol \omega}^{\hat \mu} = ({\boldsymbol
\omega}^{\hat 0}, {\boldsymbol \omega}^{\hat r}, {\boldsymbol
\omega}^{\hat \theta}, {\boldsymbol \omega}^{\hat \phi})$, whose
components are
\begin{eqnarray}
\label{tetrad_0}
&\displaystyle\boldsymbol{e}_{\hat 0}^{\alpha}=
	\frac{1}{N}\bigg(1,0,0,0\bigg) \ ,
	\hskip 4.truecm
	&\boldsymbol{\omega}^{\hat 0}_{\alpha}=
	N\bigg(1,0,0,0\bigg)\ ,             		\\
\label{tetrad_1}
&\displaystyle\boldsymbol{e}_{\hat r}^{\alpha}=
	N\bigg(0,1,0,0\bigg) \ ,
	\hskip 4.truecm
	&\boldsymbol{\omega}^{\hat r}_{\alpha}=
	\frac{1}{N}\bigg(0,1,0,0\bigg)\ ,               \\
\label{tetrad_2}
&\displaystyle\boldsymbol{e}_{\hat \theta}^{\alpha}=
	\frac{1}{r}\bigg(0,0,1,0\bigg)  \ ,
	\hskip 4.truecm
	&\boldsymbol{\omega}^{\hat \theta}_{\alpha}=
	{r}\bigg(0,0,1,0\bigg)\ ,                       \\
\label{tetrad_3}
&\displaystyle\boldsymbol{e}_{\hat \phi}^{\alpha}=
    	\frac{1}{r\sin\theta}\bigg(0,0,0,1\bigg) \ ,
    	\hskip 4.truecm
	&\boldsymbol{\omega}^{\hat \phi}_{\alpha}=
	{r\sin\theta}\bigg(0,0,0,1\bigg)\ ,
\end{eqnarray}
and which will become useful when determining the ``physical'' components
of the electromagnetic fields.

    The relations among the electric $E^{\alpha}$ and magnetic
$B^{\alpha}$ four-vector fields measured by an observer with
four-velocity $u^{\alpha}$ can be expressed through the electromagnetic
field tensor $F_{\alpha \beta}$
\begin{equation}
\label{fab_def}
F_{\alpha\beta} \equiv 2 u_{[\alpha} E_{\beta]} +
        \eta_{\alpha\beta\gamma\delta}u^\gamma B^\delta \ ,
\end{equation}
where $T_{[\alpha \beta]} \equiv \frac{1}{2}(T_{\alpha \beta} - T_{\beta
\alpha})$ and $\eta_{\alpha\beta\gamma\delta}$ is the pseudo-tensorial
expression for the Levi-Civita symbol $\epsilon_{\alpha \beta \gamma
\delta}$.

    The first pair of the general relativistic Maxwell equations
\begin{equation}
\label{maxwell_firstpair}
3! F_{[\alpha \beta, \gamma]} =  2 \left(F_{\alpha \beta, \gamma }
        + F_{\gamma \alpha, \beta} + F_{\beta \gamma,\alpha}
        \right) = 0 \ ,
\end{equation}
can then be projected along the tetrad four-vectors carried by the proper
observers and expressed in terms of the ``physical'' electric and
magnetic three-vectors ${\vec E}$ and ${\vec B}$ (Thorne et al.  1986) as
\begin{equation}
\label{max1_membrane}
{\vec \nabla} \cdot {\vec B} = 0  \ , 
	\hskip 3.0 cm 
	\partial_t {\vec B} = - {\vec \nabla} \times (N {\vec E}) \ ,
\end{equation}
where ${\vec \nabla}$ represents the covariant derivative with respect to
spatial part of the metric (\ref{schw}). Note that equations
(\ref{max1_membrane}) are not valid in a spacetime admitting a rotational
Killing vector ({\it e.g.} such as the one produced by a rotating
relativistic star) and the general form they assume in that case is given
in Appendix \ref{Lie_derivative}.

	Once expressed in their ``hatted'' component-form, equations
(\ref{max1_membrane}) simply become
\begin{eqnarray}
\label{max1a}
\sin\theta \partial_{r}\left(r^2B^{\hat r}\right)&+&
        \frac{r}{N}\partial_{\theta}\left(\sin\theta
    B^{\hat \theta}\right) +
        \frac{r}{N}\partial_{\phi} B^{\hat \phi} = 0 \ ,
\\
\label{max1b}
\left(\frac{r\sin\theta}{N}\right)\partial_t B^{\hat r}
        & = & \partial_{\phi} E^{\hat\theta}-
  \partial_{\theta}\left(\sin\theta E^{\hat \phi} \right) \ ,
\\
\label{max1c}
\left({\frac{r\sin\theta}{N}}\right)
        \partial_t B^{\hat \theta}
        &=& - \partial_{\phi}E^{\hat r} +
        \sin\theta \partial_{r}\left(r N E^{\hat \phi} \right)
        \ ,
\\
\label{max1d}
\left({\frac{r}{N} }\right)
        \partial_t B^{\hat \phi}
        &=& - \partial_{r}\left(r N E^{\hat \theta}\right)
        + \partial_{\theta}E^{\hat r} \ .
\end{eqnarray}

    Indicating now with ${\boldsymbol w}$ the conductor four-velocity and
with $\rho_e$ the proper charge density, $\rho_e w^\alpha$ will represent
the {\it convection} current, and the second pair of Maxwell equations is
then written as
\begin{equation}
\label{maxwell_secondpair}
F^{\alpha \beta}_{\ \ \ \ ;\beta} = 4 \pi J^{\alpha} =
    4\pi (\rho_e w^\alpha + j^\alpha) \ ,
\end{equation}
with the semicolon indicating the covariant derivative with respect to
the metric (\ref{schw}) and ${\boldsymbol J}$ being the {\it total}
electric-charge current. Also these equations can be written in terms of
the physical electric and magnetic vectors as
\begin{equation}
\label{max2_membrane}
{\vec \nabla} \cdot {\vec E} = 4\pi \rho_e \ , \hskip 3.0 cm \partial_t
	{\vec E} + 4\pi {\vec J} = -{\vec\nabla} \times (N {\vec B}) \ ,
\end{equation}
or, in component form, as 
\begin{eqnarray}
\label{max2a}
N \sin\theta\partial_{r}\left(r^2 E^{\hat r} \right)+
        {r}\partial_{\theta}\left(\sin\theta E^{\hat \theta}\right)
        + r \partial_{\phi}E^{\hat \phi}
        & = & {4\pi\rho_e}r^2\sin\theta \ ,
\\
\label{max2b}
\partial_{\theta}\left(\sin\theta  B^{\hat \phi} \right)
        - \partial_{\phi}B^{\hat\theta}
        & = & \left(\frac{r\sin\theta}{N}\right)
        \partial_{t} E^{\hat r} +
        {4\pi}r\sin\theta J^{\hat r} \ ,
\\
\label{max2c}
\partial_\phi B^{\hat r} - \sin\theta \partial_{r}\left(rN
        B^{\hat \phi} \right)
        & = & \left(\frac{r\sin\theta}{N}\right)
        \partial_{t} E^{\hat\theta}
        + {4\pi}r\sin\theta J^{\hat\theta} \ ,
\\
\label{max2d}
\partial_{r} \left(rN B^{\hat \theta} \right) -
        \partial_{\theta}B^{\hat r}
        & = & \left(\frac{r}{N} \right)
        \partial_{t} E^{\hat\phi}
        + {4\pi}rJ^{\hat\phi} \ .
\end{eqnarray}
Note that the total electric-charge current in equations
(\ref{maxwell_secondpair}) includes the {\it conduction} current
$j^\alpha$, which is associated with electrons having electrical
conductivity $\sigma$, and that Ohm's law can be written as
\begin{equation}
\label{ohm}
j_\alpha = \sigma F_{ \alpha \beta}w^\beta \ .
\end{equation}

    So far, we have considered the star to be static and in equilibrium,
but we can now introduce a fluid perturbation in terms of a 4-velocity
$\delta {\boldsymbol u}$, so that the fluid velocity is in general
${\boldsymbol w} \equiv \delta {\boldsymbol u}$, where the $\delta$
indicates that this is an Eulerian perturbation. The components of the
velocity perturbation are then
\begin{equation}
\label{vel}
\delta u^\alpha = \Gamma \bigg(1, \delta v^i\bigg)=
        \Gamma \bigg(1,e^{-\Lambda}\delta v^{\hat r},
	\frac{\delta v^{\hat\theta}}{r},
        \frac{\delta v^{\hat\phi}}{r\sin\theta}\bigg) \ ,
\end{equation}
and
\begin{equation}
\delta u_\alpha =
        \Gamma \bigg(-e^{2\Phi},e^{\Lambda}\delta v^{\hat r},
	{r\delta v^{\hat\theta}},
        {r\sin\theta \delta v^{\hat\phi}}\bigg) \ ,
\end{equation}
where $\delta v^i\equiv dx^i/dt$ is the oscillation 3-velocity of the
conducting stellar medium and $\delta v^{\hat i}\equiv \boldsymbol{\tilde
\omega}^{\hat i}_{k}\delta v^k$ are the components of the oscillation
3-velocity in the orthonormal frame carried by the static observers in
the stellar interior\footnote{We indicate with a tilde the 1-form basis
of the static observers inside the star to distinguish this from the
corresponding basis given in (\ref{tetrad_0}).}. Because we are
interested in small velocity perturbations for which $\delta v^i \ll 1$,
we can neglect terms ${\cal O}(\delta v^2)$ and use the normalization for
the four-velocity $w_{\alpha} w^{\alpha} = -1$ to obtain
\begin{equation}
\Gamma = \left[-g_{00}\left(1 + 
	g_{ik} \frac{\delta v^i \delta v^k}{g_{00}}\right)\right]^{-1/2} 
	\simeq e^{-\Phi}\ .
\end{equation}

	Note that $\delta {\boldsymbol u}$ is the only perturbation which
we need to consider here. While, in fact, other perturbations ({\it e.g.}
in pressure and in energy density) may be present and influence the
stellar structure, these will not affect the induction equation which
will depend linearly on $\delta {\boldsymbol u}$ only [{\it cf.} equations
(\ref{fab_def})]. Furthermore, for simplicity we will assume that $\delta
{\boldsymbol u}$ is a given, generic function and is not necessarily the
solution of an eigenvalue problem (this was considered by Messios et
al. 2001).  Finally, because we treat the electromagnetic fields as
``test-fields'', we will not consider any perturbation to the spacetime
metric (\ref{schw}), and therefore work within the so-called Cowling
approximation.

	Note that the introduction of perturbations in the velocity will
produce important modifications in the electric and magnetic fields in
three different spatial regions: {\it i)} the stellar {\it ``interior''},
$r< R$; {\it ii)} the {\it ``near-zone''}, {\it i.e.} a vacuum region
outside the stellar surface with radial extension $ r \gtrsim R$ and {\it
iii)} the {\it ``wave-zone''}, {\it i.e.} far away from the star, at $r
\gg R$, where the perturbations propagate as electromagnetic waves. Each
of these modifications has observable consequences and the following
Sections are dedicated to discussing two of these regions for the general
case of arbitrary velocity perturbations and generic magnetic field
topology.

\section{Electromagnetic Fields in the Stellar Interior}
\label{em_si}

	We here concentrate on the modifications in the {\it internal}
electromagnetic fields produced by velocity perturbations of different
types. While not directly observable, these electromagnetic fields, and
in particular the values they assume on the stellar surface, leave an
important imprint on the electromagnetic waves produced by the
oscillations and that will reach infinity.

	We here start by using Ohm's law (\ref{ohm}) and the
four-velocity (\ref{vel}), to obtain the explicit components of the total
current $J^{\hat \alpha}$ as
\begin{eqnarray}
\label{current2}
&&J^{\hat r} = {\rho_e}e^{-\Phi}\delta v^{\hat r}+ \sigma
        \left[ E^{\hat r} +e^{-\Phi}\left(\delta v^{\hat\theta} B^{\hat\phi}
        -\delta v^{\hat\phi}B^{\hat \theta}\right)\right] \ ,
\\\nonumber\\
\label{current3}
&&J^{\hat\theta} = {\rho_e}e^{-\Phi}{\delta v^{\hat \theta}}
    + \sigma\left[E^{\hat \theta}+
        e^{-\Phi}\left(\delta v^{\hat\phi} B^{\hat r}
        -\delta v^{\hat r}B^{\hat \phi}\right)\right] \ ,
\\\nonumber\\
\label{current4}
&&J^{\hat\phi} = {\rho_e}e^{-\Phi}{\delta v^{\hat \phi}}
    + \sigma \left[E^{\hat \phi} +
        e^{-\Phi}\left(\delta v^{\hat r} B^{\hat\theta}
        -\delta v^{\hat \theta}B^{\hat r}\right)\right] \ .
\end{eqnarray}

	A convenient way of simplifying the problem is that of
considering the fluid as perfectly conducting, {\it i.e.} with $\sigma
\rightarrow \infty$ (this is the limit of ideal
magnetohydrodynamics). While idealized, this approximation is a rather
good one in the present case, since the Ohmic diffusion timescale is
several orders of magnitude larger that the typical timescale for the
stellar oscillations.  Once this assumption is made, the electric field
in the interior of the star can be easily derived from (\ref{current2})--
(\ref{current4}) to be
\begin{eqnarray}
\label{in_ef2}
&&E_{\rm int}^{\hat r} =
        -e^{-\Phi}\left(\delta v^{\hat\theta} B_{\rm int}^{\hat\phi}
        -\delta v^{\hat\phi}B_{\rm int}^{\hat \theta}\right) \ ,
\\\nonumber\\
\label{in_ef3}
&&E^{\hat\theta}_{\rm int} =
-       e^{-\Phi}\left(\delta v^{\hat\phi} B_{\rm int}^{\hat r}
        -\delta v^{\hat r}B_{\rm int}^{\hat \phi}\right) \ ,
\\\nonumber\\
\label{in_ef4}
&&E_{\rm int}^{\hat\phi} =
-       e^{-\Phi}\left(\delta v^{\hat r} B_{\rm int}^{\hat\theta}
        -\delta v^{\hat \theta}B_{\rm int}^{\hat r}\right) \ ,
\end{eqnarray}
where we have indicated with an index ``int'' the field components for
$r\le R$. The Maxwell equations (\ref{max1a})--(\ref{max1d}) combined
with the expressions ({\ref{in_ef2})--(\ref{in_ef4}) will lead to the
following set of the magnetic induction equations 
\begin{eqnarray}
\label{ind1}
\left({r\sin\theta}\right)\partial_t B_{\rm int}^{\hat r}
        & = & \partial_\phi \left(\delta v^{\hat r} B_{\rm int}^{\hat\phi}
        -\delta v^{\hat\phi}B_{\rm int}^{\hat r}\right)
        +\partial_\theta\left[\sin\theta
         \left(\delta v^{\hat r} B_{\rm int}^{\hat\theta}
        -\delta v^{\hat\theta}B_{\rm int}^{\hat r}\right)\right]\ ,
\\\nonumber\\
\label{ind2}
\left(r\sin\theta\right)
        \partial_t B_{\rm int}^{\hat \theta}
        &=& \partial_\phi \left(\delta v^{\hat\theta} B_{\rm int}^{\hat\phi}
        -\delta v^{\hat\phi}B_{\rm int}^{\hat\theta}\right)
        -\sin\theta e^{-\Lambda}\partial_r\left[r
         \left(\delta v^{\hat r} B_{\rm int}^{\hat\theta}
        -\delta v^{\hat\theta}B_{\rm int}^{\hat r}\right)\right]\ ,
\\\nonumber\\
\label{ind3}
r\partial_t B_{\rm int}^{\hat \phi}
        &=& e^{-\Lambda}\partial_r\left[r \left(\delta v^{\hat\phi}
        B_{\rm int}^{\hat r}
        -\delta v^{\hat r}B_{\rm int}^{\hat\phi}\right)\right]
        -\partial_\theta
         \left(\delta v^{\hat\theta} B_{\rm int}^{\hat\phi}
        -\delta v^{\hat\phi}B_{\rm int}^{\hat\theta}\right) \ .
\end{eqnarray}

	The boundary conditions for the magnetic field across the stellar
surface $r=R$ can then be obtained after requiring continuity for the
normal ({\it i.e.} $r$) component
\begin{eqnarray}
\label{bound_b1}
&&B^{\hat r}_{\rm ext}|_{r=R}=B^{\hat r}_{_R}\ ,
\end{eqnarray}
while leaving the tangential ({\it i.e.} $\theta$ and $\phi$) components
free to be discontinuous through surface currents
\begin{eqnarray}
\label{bound_b2}
&&B^{\hat\theta}_{\rm ext}|_{r=R}=B^{\hat\theta}_{_R}
    +4\pi I^{\hat\phi}\ ,
\\\nonumber\\
\label{bound_b3}
&&B^{\hat\phi}_{\rm ext}|_{r=R}=B^{\hat\phi}_{_R}
    -4\pi I^{\hat\theta}\ ,
\end{eqnarray}
where $I^{\hat k}$ are the components of the surface current and $B^{\hat
i}_{_R} \equiv B^{\hat i}_{\rm int}|_{r=R}$ (the index ``ext'' has been
here used to refer to fields at $r\ge R$). In a similar way, the boundary
conditions for the electric field across the stellar surface can be
derived from imposing the continuity of the tangential components, while
leaving the normal one free to be discontinuous through a surface charge
distribution $\Sigma_s$. Simple algebra then gives
\begin{eqnarray}
\label{bound_e1}
&&E^{\hat r}_{\rm ext}|_{r=R}=E^{\hat r}_{\rm int}|_{r=R}
    +4\pi\Sigma_s=
        -\frac{1}{N_{_R}}\left(\delta v_{_R}^{\hat\theta} B^{\hat\phi}_{_R}
        -\delta v_{_R}^{\hat\phi}B^{\hat \theta}_{_R}\right) +4\pi\Sigma_s \ ,
\\\nonumber\\
\label{bound_e2}
&&E^{\hat\theta}_{\rm ext}|_{r=R}=E^{\hat\theta}_{\rm int}|_{r=R}=
    -\frac{1}{N_{_R}}\left(\delta v_{_R}^{\hat\phi} B^{\hat r}_{_R}
        -\delta v_{_R}^{\hat r}B^{\hat \phi}_{_R}\right) \ ,
\\\nonumber\\
\label{bound_e3}
&&E^{\hat\phi}_{\rm ext}|_{r=R}=E^{\hat\phi}_{\rm int}|_{r=R}=
    -\frac{1}{N_{_R}}\left(\delta v_{_R}^{\hat r} B^{\hat\theta}_{_R}
        -\delta v_{_R}^{\hat \theta}B^{\hat r}_{_R}\right) \ ,
\end{eqnarray}
where $\delta v^{\hat i}_{_R} \equiv \delta v^{\hat i}|_{r=R}$ are the
oscillation velocities at the stellar surface, the only relevant ones in
this paper.

\section{Electromagnetic Fields in the Near-Zone}
\label{em_nz}

	We now consider the form of the electromagnetic fields generated
by stellar oscillations in a region of vacuum space exterior and close to
the stellar surface. Our treatment in this Section will be as general as
possible and we will not restrict it to any specific magnetic field
topology or velocity field. A more detailed discussion of the
electromagnetic fields produced by those spheroidal and toroidal velocity
fields most often discussed in the physics of compact stars will be
presented in a companion paper \citep{r3}.

	Given a characteristic frequency for the stellar oscillations
$f_0$, the spatial extension of the {\it near-zone} can then be estimated
to be $\lambda \sim c/f_0$, which is the distance travelled by an
electromagnetic wave in a timescale $1/f_0$. For a typical neutron star
of mass $M$ and radius $R$, the fundamental frequency of oscillation is
$f_0 \sim \sqrt{GM/R^3} \sim 2-3$ kHz, so that the near-zone extends from
$R$ to $\sim 100\ R$. Since we are neglecting terms ${\cal O}(\delta
v^2)$ in the Maxwell equations, we need not account for the displacement
currents in equations (\ref{max2b})--(\ref{max2d}) and set
\begin{equation}
\label{no_disp_currnt}
        \partial_{t} E^{\hat r}=
        \partial_{t} E^{\hat\theta}=
        \partial_{t} E^{\hat\phi}\equiv 0 \ ,
\end{equation}
so that the second pair of Maxwell equations can be written as
\begin{eqnarray}
\label{max_near_2a}
N \sin\theta\partial_{r}\left(r^2 E^{\hat r} \right)+
        {r}\partial_{\theta}\left(\sin\theta E^{\hat \theta}\right)
        + r \partial_{\phi}E^{\hat \phi}
        = 0 \ ,&&
\\
\label{max_near_2b}
\partial_{\theta}\left(\sin\theta  B^{\hat \phi} \right)
        - \partial_{\phi}B^{\hat\theta}
        = 0 \ ,&&
\\
\label{max_near_2c}
\partial_{\phi}B^{\hat r} - \sin\theta \partial_{r}\left(rN
        B^{\hat \phi} \right)
        = 0 \ ,&&
\\
\label{max_near_2d}
\partial_{r}\left(Nr B^{\hat \theta} \right) -
        \partial_{\theta}B^{\hat r}
        = 0 \ .&&
\end{eqnarray}

    The solutions to the Maxwell equations with sources represented by a
bounded distribution of currents that vary harmonically with time is
mostly easily found if the electromagnetic fields are expanded in their
multipolar components. An efficient method to obtain this multipolar
expansions has been proposed by Bouwkamp \& Casimir (1954) in Newtonian
electrodynamics and uses the general properties of solenoidal vector
fields. We here follow the same approach but extend it to relativistic
electrodynamics. Consider therefore four scalar functions $S, T, X$, and
$Z$, through which the general solutions to the vacuum Maxwell equations
(\ref{max1a})--(\ref{max1d}) and (\ref{max_near_2a})--(\ref{max_near_2d})
can be written in the form
\begin{eqnarray}
\label{sol_mf1}
&& B^{\hat r} = -\frac{1}{r^2\sin^2\theta}\left[
    \sin\theta\partial_{\theta}\left(\sin\theta
    \partial_{\theta}S\right)+\partial^2_{\phi}S\right]\ ,
\\\nonumber\\
\label{sol_mf2}
&& B^{\hat\theta} =\frac{N}{r}\partial_{\theta}\partial_{r}S
    + \frac{1}{Nr\sin\theta}\partial_{\phi}Z\ ,
\\\nonumber\\
\label{sol_mf3}
&& B^{\hat\phi} = \frac{N}{r\sin\theta}\partial_{\phi}\partial_{r}S
    - \frac{1}{Nr\sin\theta}\partial_{\theta}Z \ ,
\end{eqnarray}
and
\begin{eqnarray}
\label{sol_ef1}
&& E^{\hat r} = -\frac{1}{r^2\sin^2\theta}\left[
    \sin\theta\partial_{\theta}\left(\sin\theta
    \partial_{\theta}T\right)+\partial^2_{\phi}T\right]\ ,
\\\nonumber\\
\label{sol_ef2}
&& E^{\hat\theta} =\frac{N}{r}\partial_{\theta} \partial_{r}T
    + \frac{1}{Nr\sin\theta}\partial_{\phi}X\ ,
\\\nonumber\\
\label{sol_ef3}
&& E^{\hat\phi} = \frac{N}{r\sin\theta}\partial_{\phi} \partial_{r}T
    - \frac{1}{Nr}\partial_{\theta}X \ .
\end{eqnarray}
We will refer to the functions $S, X$ and $T, Z$ as to the ``magnetic''
and ``electric'' functions, respectively. Furthermore, to handle
analytically the angular derivatives, we will assume they can be
separated in variables and expand the angular dependence in terms of
spherical harmonics $Y_{\ell m}(\theta,\phi)$
\begin{eqnarray}
&\displaystyle{S(t,r,\theta ,\phi)=
	\sum^{\infty}_{\ell=0}\sum^{\ell}_{m=-\ell}
	{S}_{\ell m}(t,r) Y_{\ell m}(\theta,\phi)} \ ,
\hskip 2.0 cm 
&X(t,r,\theta ,\phi)=
	\sum^{\infty}_{\ell=0}\sum^{\ell}_{m=-\ell}{X}_{\ell m}(t,r)
    Y_{\ell m}(\theta,\phi) \ ,
\\
\hskip 0.01 cm 
&\displaystyle{T(t,r,\theta ,\phi)=
	\sum^{\infty}_{\ell=0}\sum^{\ell}_{m=-\ell}
	{T}_{\ell m}(t,r)Y_{\ell m}(\theta,\phi)} \ ,
\hskip 2.0 cm 
&Z(t,r,\theta ,\phi)=
	\sum^{\infty}_{\ell=0}\sum^{\ell}_{m=-\ell}{Z}_{\ell m}(t,r)
	Y_{\ell m}(\theta,\phi) \ .
\end{eqnarray}

	Using this {\sl ansatz} and suppressing the summation symbols
over the $\ell,m$ indices, the general solutions
(\ref{sol_mf1})--(\ref{sol_ef3}) then take the form
\begin{eqnarray}
\label{sol_s_mf1}
&& B^{\hat r} = \frac{\ell\left(\ell+1\right)}{r^2}
    S_{\ell m}Y_{\ell m}\ ,
\\\nonumber\\
\label{sol_s_mf2}
&& B^{\hat\theta} =\frac{N}{r}\partial_{r}S_{\ell m}
        \partial_{\theta}Y_{\ell m}
    + \frac{1}{Nr\sin\theta}Z_{\ell m}\partial_{\phi}Y_{\ell m}\ ,
\\\nonumber\\
\label{sol_s_mf3}
&& B^{\hat\phi} = \frac{N}{r\sin\theta}\partial_{r}S_{\ell m}
    \partial_{\phi}Y_{\ell m} -
        \frac{1}{Nr}Z_{\ell m}\partial_{\theta}Y_{\ell m} \ ,
\end{eqnarray}
for the magnetic field components, and
\begin{eqnarray}
\label{sol_s_ef1}
&& E^{\hat r} = \frac{\ell\left(\ell+1\right)}{r^2}
    T_{\ell m}Y_{\ell m}\ ,
\\\nonumber\\
\label{sol_s_ef2}
&& E^{\hat\theta} =\frac{N}{r}\partial_{r}T_{\ell m}\partial_{\theta}Y_{\ell m}
    + \frac{1}{Nr\sin\theta}X_{\ell m}\partial_{\phi}Y_{\ell m}\ ,
\\\nonumber\\
\label{sol_s_ef3}
&& E^{\hat\phi} = \frac{N}{r\sin\theta}\partial_{r}T_{\ell m}
    \partial_{\phi}Y_{\ell m} -
        \frac{1}{Nr}X_{\ell m}\partial_{\theta}Y_{\ell m} \ ,
\end{eqnarray}
for the electric field components. We next concentrate on how to use the
Maxwell equations to recast expressions
(\ref{sol_s_mf1})--(\ref{sol_s_ef3}) in a form more useful for
astrophysical applications.

\subsection{Magnetic Fields}

    Substituting the general solutions
(\ref{sol_s_mf1})--(\ref{sol_s_mf3}) for the magnetic field into the
Maxwell equations (\ref{max_near_2b})--(\ref{max_near_2d}) gives two
equations for the unknown functions $Z_{\ell m}$ and $S_{\ell m}$
\begin{eqnarray}
\label{eqn_Z}
\left[\frac{1}{\sin\theta}\partial_{\theta}\left(\sin\theta
    \partial_{\theta}Y_{\ell m}\right)+
    \frac{1}{\sin^2\theta}\partial^2_{\phi}Y_{{\ell m}}\right]
    Z_{\ell m}=0 \ ,
\\\nonumber\\
\label{eqn_S}
\frac{d}{dr}\left[\left(1-\frac{2M}{r}\right)
    \frac{d}{dr} {S}_{\ell m}\right]
    -\frac{\ell\left(\ell+1\right)}{r^2}{S}_{\ell m}=0\ ,
\end{eqnarray}
which have the important property of depending on $Z$ or $S$ only, so
that they can be solved separately. The first equation has indeed the
trivial solution $Z_{\ell m} = 0$ since the content of the square
brackets in (\ref{eqn_Z}) is a function of $\theta$ and $\phi$ only. The
solution of (\ref{eqn_S}), on the other hand, is less straightforward. It
is convenient to introduce the new variable
\begin{equation}
\label{def_x}
x\equiv 1-\frac{r}{M}\ ,
\end{equation}
and factor out an explicit quadratic radial dependence in the functions
$S_{\ell m}$ which are then redefined as
\begin{equation}
\label{def_slm}
{S}_{\ell m}\equiv r^2 h(t,r) \ .
\end{equation}
Using now (\ref{def_x}) and (\ref{def_slm}), equation (\ref{eqn_S})
takes the form
\begin{equation}
\label{eqn_P}
\frac{d}{dx}\left\{\left(\frac{1+x}{1-x}\right)
    \frac{d}{dx}\left[\left(1-x\right)^2 h\right]\right\}
    +\ell\left(\ell+1\right)h=0\ ,
\end{equation}
which can be solved in terms of Legendre functions of the second kind
$Q_{\ell}(x)$ (see Rezzolla et al. 2001a for details) so that
\begin{equation}
\label{sol_P}
{S}_{\ell m}(t,r)= \frac{(1-x)^2}{M}
	\frac{d}{dx}\left[\left({1+x}\right)
	\frac{dQ_{\ell}}{dx}\right] s_{\ell m}(t)\ .
\end{equation}
Note that all of the time dependence in (\ref{sol_P}) is contained in the
integration constants $s_{\ell m}(t)$ and these are determined after
suitable boundary conditions across the stellar surface are imposed.

	We next consider how these expressions vary when a perturbation,
caused for instance by a nonzero velocity field (\ref{vel}), is
introduced. Hereafter, we will assume that a background magnetic field is
present and which varies with time only on a timescale which is much
longer than that of the stellar oscillations and can therefore be
considered static. Hereafter we will indicate with $B^{\hat i}_0$ the
components of this zeroth-order (background) stellar magnetic field and
with ${\delta B}^{\hat i}(t)$ the first-order perturbations, which
clearly are the ones possessing a time-dependence.

	Within a linear regime in the perturbations, a convenient way to
isolate the perturbed part of the magnetic fields is suggested by the
structure of the solutions (\ref{sol_s_mf1})--(\ref{sol_s_mf3}) which
confines all of the time dependence in the $s_{\ell m}$ terms. As a
result, it is possible to express the magnetic field perturbation in
terms of new, time-dependent integration constants $\delta s_{\ell m}(t)$
that simply add to the background ones $s_{\ell m}$. The new components
of the magnetic field generated by a velocity perturbation will therefore
have the generic form
\begin{eqnarray}
\label{nz_mfg1}
&& B^{\hat r} = 
	B^{\hat r}_0+\delta B^{\hat r}(t) = \frac{\ell(\ell+1)}{M^3}
	\frac{d}{dx}\left[\left({1+x}\right)
	\frac{dQ_{\ell}}{dx}\right]
        \left[s_{\ell m}+\delta s_{\ell m}(t)\right] Y_{\ell m}\ ,
\\ \nonumber    \\
\label{nz_mfg2}
&& B^{\hat \theta} = 
	B^{\hat \theta}_0+\delta B^{\hat \theta}(t) =
        -\left[-\frac{1+x}{(1-x)^3}\right]^{1/2}
	\frac{1}{M^3}\frac{d}{dx}\left\{ (1-x)^2\frac{d}{dx}
	\left[\left({1+x}\right)
	\frac{dQ_{\ell}}{dx}\right]  \right\}
        \left[s_{\ell m}+\delta s_{\ell m}(t)\right]
	\partial_{\theta}  Y_{\ell m} \ ,
\\ \nonumber    \\
\label{nz_mfg3}
&& B^{\hat \phi} = 
	B^{\hat \phi}_0+\delta B^{\hat \phi}(t) =
        -\left[-\frac{1+x}{(1-x)^3}\right]^{1/2}
	\frac{1}{M^3}\frac{d}{dx}\left\{ (1-x)^2 \frac{d}{dx}
	\left[\left({1+x}\right)
	\frac{dQ_{\ell}}{dx}\right]  \right\}
        \left[s_{\ell m}+\delta s_{\ell m}(t)\right]
	\frac{1}{\sin\theta}\partial_{\phi}  Y_{\ell m}
	\ .
\end{eqnarray}

	The values of the integration constants $\delta s_{\ell m}(t)$
can be calculated rather straightforwardly if the oscillation modes are
assumed to have a simple harmonic time dependence of the type ${\rm
exp}(-i\omega t)$, where $\omega$ is the mode frequency. In this case, in
fact, the radial component for the induction equation~(\ref{ind1}) at the
surface of the star becomes
\begin{equation}
\label{delta_B}
\partial_t (\delta B_{R}^{\hat r})=-i\omega_{_{R}} \delta B_{R}^{\hat r}
         =  \frac{1}{R\sin\theta}\left\{
        \partial_\phi \left(\delta v_{_R}^{\hat r} B_{R}^{\hat\phi}
        -\delta v_{_R}^{\hat\phi}B_{R}^{\hat r}\right)
        + \partial_\theta\left[\sin\theta
         \left(\delta v_{_R}^{\hat r} B_{R}^{\hat\theta}
        -\delta v_{_R}^{\hat\theta}B_{R}^{\hat r}
	\right)\right]\right\}\ ,
\end{equation}
and the boundary condition for the continuity of the radial magnetic
field (\ref{bound_b1}) can then be used to determine the coefficients
$\delta s_{\ell m}(t)$ (see Appendix~\ref{der_t_delta_s} for details).
Note that this procedure is different from the one used by Timokhin et
al., (2000), in which the integration constants are instead obtained
through the boundary conditions applied to the continuity of the
tangential components of the electric field. Of course, the two
procedures are equivalent and yield identical results.

	A few remarks are worth making at this point. The first one is
about the assumption of infinite conductivity for the stellar material,
which implies that the magnetic field is advected with the fluid when
this possesses a nonzero velocity. This is expressed by the general
relativistic ``frozen-flux'' condition\footnote{Note that the lapse
function does not appear multiplying the magnetic field in expression
(\ref{ff}). This is because of a cancellation when Ohm's law is used to
replace the contribution to the Maxwell equations coming from the
electric field (see Appendix \ref{Lie_derivative} for the expression of
Ohm's law in terms of physical three-vectors).}
\begin{equation}
\label{ff}
\partial_t \delta {\vec B} = {\vec \nabla} \times \left(\delta {\vec v} 
	\times {\vec B}_0\right) \ ,
\end{equation}
and is, indeed, verified by expressions
(\ref{nz_mfg1})--(\ref{nz_mfg3}). The second one is that these
expressions, although interesting for their generality, are not
particularly useful if we do not restrict our attention to a specific
background magnetic field configuration; this will be done in
Section~\ref{dipolarmf} where we concentrate on the specific form that
equations (\ref{nz_mfg1})--(\ref{nz_mfg3}) assume for a background static
dipolar magnetic field. Finally, note that no reference is made at this
point on the type of velocity perturbations introduced as these are
effectively incorporated in the expressions for the functions $\delta
s_{\ell m}(t)$; in Sections~\ref{rd_nz} and \ref{rd_wz} we will discuss
the form of these functions for the case of a magnetic dipole in uniform
rotation.

\subsection{Electric Fields}

	We next concentrate on the form of the electric field components
and consider for this the solution to the Maxwell equations
(\ref{max1b})--(\ref{max1d}) and (\ref{max_near_2a}). Hereafter we will
assume that the star has a zero net electric charge ({\it i.e.} $\int
\rho_e \sqrt{\gamma} d^3{\bf x} =0$, where $\gamma$ is the determinant of
the three-metric and $d^3{\bf x}$ is the coordinate volume element) and
that the background magnetic field is stationary ({\it i.e.} $\partial_t
B^{\hat i}_0 =0$). In this case, the background electric fields will be
identically zero ({\it i.e.} $E^{\hat i}_0=0$) but time-dependent,
first-order electric fields $\delta E^{\hat i}(t)$ may be present and
induced by the perturbations in the background magnetic field.

	To derive the expressions for these perturbed electric fields we
substitute the decomposed expressions (\ref{sol_ef1})--(\ref{sol_ef3})
into the Maxwell equations (\ref{max1b})--(\ref{max1d}) and
(\ref{max_near_2a}) to obtain the following set of equations for the
unknown functions $T_{\ell m}$ and $X_{\ell m}$
\begin{eqnarray}
\label{eqn_T}
&& \frac{d}{dr}\left[\left(1-\frac{2M}{r}\right)
    \frac{d}{dr} {T}_{\ell m}\right]
    -\frac{\ell\left(\ell+1\right)}{r^2}{T}_{\ell m}=0\ ,
\\\nonumber\\
\label{eqn_X}
&& \ell(\ell+1)Y_{\ell m}X_{\ell m}=-r^2\partial_t B^{\hat r}\ ,
\\\nonumber\\
&& \partial_{r} X_{\ell m}\partial_{\theta}Y_{\ell m}=
-\frac{r}{N}\partial_t B^{\hat\theta} \ ,
\\\nonumber\\
&& \partial_r X_{\ell m}\partial_{\phi}Y_{\ell m}=
-\frac{r\sin\theta}{N}\partial_t B^{\hat\phi} \ .
\end{eqnarray}

	It is apparent that equation (\ref{eqn_T}) is the same as
(\ref{eqn_S}) but for the unknown functions ${T}_{\ell m}$ and will
therefore have solutions in terms of Legendre functions of the second
kind $Q_{\ell}(x)$ and the integration constants $\delta t_{\ell m}(t)$
\begin{equation}
\label{sol_T}
{T}_{\ell m} =
	(1-x)^2 M^2 \frac{d}{dx}\left[\left({1+x}\right)
	\frac{dQ_{\ell}}{dx}\right] \delta t_{\ell m}(t)\ .
\end{equation}
In general, that is for a multipolar magnetic field, the ``electric''
functions $X_{\ell m}(t)$ are determined directly from the time
derivatives of the ``magnetic'' functions $S_{\ell m}$
[{\it cf.} eqs. (\ref{sol_s_mf1}) and (\ref{eqn_X})], {\it i.e.}
\begin{equation}
\label{sol_X}
X_{\ell m}=
	-\partial_t S_{\ell m}=
	\frac{(1-x)^2}{M}\frac{d}{dx}\left[\left({1+x}\right)
	\frac{dQ_{\ell}}{dx}\right] \delta x_{\ell m}(t)\ ,
\end{equation}
where $\delta x_{\ell m}(t) = - \partial_t[\delta s_{\ell
m}(t)]$. Expression (\ref{sol_X}) underlines that the functions $X_{\ell
m}$ represent the contribution to the electric field coming from the time
variation of the magnetic field, with the coefficients $\delta x_{\ell
m}$ being determined, also in this case, through the use of suitable
boundary conditions.

	Using expressions (\ref{sol_T}) and (\ref{sol_X}) and standard
recurrence formulae for the Legendre functions that we will recall in
Appendix A, the solution for the vacuum electric field
(\ref{sol_s_ef1})--(\ref{sol_s_ef3}) can be written as
\begin{eqnarray}
\label{nz_ef1}
&& \delta E^{\hat r} = 
	\frac{\ell^2\left(\ell+1\right)}{(1-x)^2}
	\left[Q_{\ell -1} -\left(\ell + 1 - \ell x \right)
	Q_{\ell} \right]
	\delta t_{\ell m}(t) Y_{\ell m}\ ,
\\ \nonumber\\
\label{nz_ef2}
&& \delta E^{\hat\theta} = 
	\frac{\ell^2\left(\ell+1\right)}{(1-x)^2 N}
	\left[x	Q_{\ell} - Q_{\ell -1}\right]
	\delta t_{\ell m}(t) \partial_{\theta}Y_{\ell m}
	+ \frac{(1-x)M}{N\sin\theta}
	\frac{d}{dx}\left[\left({1+x}\right)
    	\frac{dQ_{\ell}}{dx}\right] 
	\delta x_{\ell m}(t) \partial_{\phi}Y_{\ell m} \ ,
\\ \nonumber\\
\label{nz_ef3}
&& \delta E^{\hat\phi} = 
	\frac{\ell^2\left(\ell+1\right)}{(1-x)^2 N\sin\theta}
	\left[x	Q_{\ell} - Q_{\ell -1}\right]
	\delta t_{\ell m}(t)\partial_{\phi}Y_{\ell m}
	-\frac{(1-x)M}{N}
	\frac{d}{dx}\left[\left({1+x}\right)
    	\frac{dQ_{\ell}}{dx}\right] 
	\delta x_{\ell m}(t)\partial_{\theta} Y_{\ell m} \ ,
\end{eqnarray}
where, again, $Q_{\ell} = Q_{\ell}(x)$. Note how equations
(\ref{nz_ef1})--(\ref{nz_ef3}) clearly indicate that given a magnetic
field with multipolar components up to the order $\ell$, the
corresponding electric field will have multipolar components up to the
order $\ell + 1$. This is another aspect of the interconnection between
electric and magnetic fields contained in the Maxwell equations.

	Finally, the integration constants $\delta t_{\ell m}$ and
$\delta x_{1m}$ can be calculated from the jump conditions
(\ref{bound_e2}) --(\ref{bound_e3}) of the tangential electric field
(\ref{nz_ef2})--(\ref{nz_ef3}) at the stellar surface and are given by
\begin{eqnarray}
\label{tlm}
&& \delta t_{\ell m}(t) = \frac{(1-x_{_R})^2}
	{\ell^3\left(\ell+1\right)^2}
    	\left[x_{_R}
	Q_{\ell}(x_{_R}) - Q_{\ell -1}(x_{_R})\right]^{-1} \times
\nonumber \\
&& \hskip 5.0 cm 	
    \times \int d\Omega\left\{\partial_{\theta}Y^*_{\ell m}
        \left[\delta v_{_R}^{\hat\phi}(t) B^{\hat r}_{_R}
        -\delta v_{_R}^{\hat r}(t) B^{\hat\phi}_{_R}\right]
        -i\frac{mY^*_{\ell m}}{\sin\theta}
        \left[\delta v_{_R}^{\hat\theta}(t) B^{\hat r}_{_R}
        -\delta v_{_R}^{\hat r}(t) B^{\hat\theta}_{_R}\right]\right\} \ ,
\\\nonumber \\
\label{xlm}
&& \delta x_{\ell m}(t) = - \frac{3}{8}\frac{(1-x_{_R})^2}{M f_{_R}}
    \int d\Omega\left\{\partial_{\theta}Y^*_{\ell m}
        \left[\delta v_{_R}^{\hat\theta}(t) B^{\hat r}_{_R}
        -\delta v_{_R}^{\hat r}(t) B^{\hat\theta}_{_R}\right]
        +\frac{mY^*_{\ell m}}{\sin\theta}
        \left[\delta v_{_R}^{\hat\phi}(t) B^{\hat r}_{_R}
        -\delta v_{_R}^{\hat r}(t) B^{\hat\phi}_{_R}\right]\right\} \ ,
\end{eqnarray}
where $Q_{\ell}(x_{_R}) \equiv Q_{\ell}(1-R/M)$, $d\Omega =\sin\theta
d\theta d\phi$, and $f_{_R}$ is just shorthand notation for
\begin{eqnarray}
\label{f_R}
&& f_{_R} \equiv
	-\frac{3}{8}\left(\frac{R}{M}\right)^3\left[
	\ln N_{_R}^2+\frac{2M}{R}\left(1+\frac{M}{R}\right)  
	\right] =
	-\frac{3}{8}\left(\frac{R}{M}\right)^3\left[
	\ln \left(1-\frac{2M}{R}\right)+\frac{2M}{R}
	\left(1+\frac{M}{R}\right) 
	\right]
	\ .
\end{eqnarray}

\section{Electromagnetic Fields in the Wave-Zone}
\label{em_wz}

	The treatment of the electromagnetic fields in the wave-zone can
start from considering the Maxwell equations (\ref{max1a})--(\ref{max1d})
and (\ref{max2a})--(\ref{max2d}) in the case when the electric charges
and currents are absent. In particular, inserting equations (\ref{max1c})
and (\ref{max1d}) in the time derivative of equation (\ref{max2b}) and
using the expression for (\ref{max2a}) in vacuum ({\it i.e.} for
$\rho_e=0$), one obtains the following wave equation for the radial
component of electric field on a curved Schwarzschild background
\begin{equation}
\label{wave_eqn}
\partial^2_{t}E^{\hat r} - \frac{N^2}{r^2}\partial_{r}\left[N^2\partial_{r}
    \left(r^2E^{\hat r}\right)\right] -
    \frac{N^2}{r^2\sin\theta}
    \partial_{\theta}\left(\sin\theta \partial_{\theta}E^{\hat r}\right)-
    \frac{N^2}{r^2\sin^2\theta}\partial^2_{\phi}E^{\hat r}
    =0 \ .
\end{equation}

	Substituting the decomposition (\ref{sol_s_ef1}) for the radial
component of the electric field in terms of spherical harmonics, into the
wave equation (\ref{wave_eqn}) yields the well-known Regge-Wheeler
equation describing the dynamics of vector perturbations in a
Schwarzschild spacetime (Chandrasekhar, 1992)
\begin{equation}
\label{regge}
\left(\partial_{t}^2 - \partial^2_{r_*}\right)T_{\ell m} +
	\ell(\ell+1)\frac{N^2}{r^2}T_{\ell m} = 0\ ,
\end{equation}
where $r_*\equiv r+2M \ln\left({r}/{2M}-1\right)$ is the tortoise
coordinate (Similar equations can be obtained also for the other
functions $S_{\ell m}, X_{\ell m}$, and $Z_{\ell m}$.).

	Although equation (\ref{regge}) has been extensively studied in
the past and contains information about general relativistic effects such
as the scattering off the curved background spacetime (see Malec et
al. 1998 for a recent investigation), it does not represent the most
convenient way to evaluate the components of the magnetic and electric
fields in the wave-zone. There, in fact, the relativistic corrections
${\cal O}(M/r)$ can be reasonably neglected and we can seek guidance in
determining the components of the electromagnetic fields from simpler
Newtonian expressions. 

	We note that assuming a flat background spacetime for the form of
the electromagnetic fields in the wave-zone is clearly an approximation,
but it is not in contrast with the general relativistic approach followed
so far. As will become clearer in the following, in fact, the final
expressions which we derive on a flat background, still need to be
completed through the specification of suitable boundary conditions to be
imposed at the stellar surface and these will indeed contain the general
relativistic character of the curved spacetime. Furthermore, adopting a
Newtonian framework as a guideline is not only a good approximation, but
also has two important advantages. The first one is that is possible to
find analytic expression to the electromagnetic fields which will depend
on outgoing spherical Hankel functions and their derivatives
[{\it cf.} equations (\ref{sol_wz_mf1})--(\ref{sol_wz_ef3})]. In a fully
general relativistic treatment, instead, the numerical solution of an
ordinary differential equation is necessary to obtain the form of the
electromagnetic fields in the wave-zone (this will be discussed in
Rezzolla \& Ahmedov 2004). The second advantage is that in the specific
but important application of the present formalism to a rotating magnetic
dipole, the well-known Newtonian expressions can find general
relativistic corrections that have simple physical interpretations (this
will be discussed in Section~\ref{rmd}).

	We therefore express the electromagnetic fields in the wave-zone
through two scalar functions $U$ and $V$ (also sometimes referred to as
the ``Debye potentials''), whose angular dependence is expanded in a
series of spherical harmonics and that have a harmonic time dependence
\begin{equation}
U(t,r,\theta,\phi) \equiv 
	rU_{\ell m}(r)Y_{\ell m}(\theta,\phi)e^{-i\omega t}\ ,
    \hskip 3.0 cm
V(t,r,\theta,\phi) \equiv 
	rV_{\ell m}(r)Y_{\ell m}(\theta,\phi)e^{-i\omega t}\ .
\end{equation}
Note that in the curved spacetime exterior to the star, the observed
frequency for the stellar oscillations will be different for different
observers and therefore is a function of position. In particular, if
$k^{\alpha}$ is the null wave-vector associated with the electromagnetic
fields in the wave-zone and $u^{\alpha}$ the 4-velocity of an observer,
then $\omega \equiv - k^\alpha u_\alpha$ will be the frequency measured
by such an observer. Denoting with $\omega_{_R}\equiv \omega(r=R)$ the
angular frequency of oscillation measured by an observer at the stellar
surface, the corresponding electromagnetic wave frequency at a generic
radial position $r > R$ will be subject to the standard gravitational
redshift and given by (see Appendix \ref{wave_equation} for a somewhat
different derivation)
\begin{equation}
\label{grs}
\omega(r) = \omega_{_R} \frac{N_{_R}}{N} =
    \omega_{_R} \sqrt{\left(\frac{R-2M}{r-2M}\right)\frac{r}{R}} \ .
\end{equation}
In the asymptotically flat regions of the spacetime where we are
considering the wave-zone solutions to the Maxwell equations, the
electromagnetic waves will have reached their asymptotic form and their
frequency can then be considered to be simply $\omega \simeq \omega_{_R}
N_{_R}=$ const.

	Under these assumptions, and following the approach suggested by
Bouwkamp \& Casimir (1954), and Casimir (1960), the components of the
electromagnetic fields in the wave-zone assume the generic form
\begin{eqnarray}
\label{sol_w_mf1}
&& B^{\hat r} = \frac{1}{r} U_{\ell m}Y_{\ell m}
	e^{-i\omega t}\ ,
\\\nonumber\\
\label{sol_w_mf2}
&& B^{\hat\theta}
	= \frac{1}{r\ell\left(\ell+1\right)}
	\left[\partial_{r}\left(rU_{{\ell m}}\right)
	\partial_{\theta}Y_{\ell m}
	- \frac{i\omega}{\sin\theta}rV_{\ell m}\partial_{\phi}Y_{\ell m}
	\right]e^{-i\omega t} \ ,
\\\nonumber\\
\label{sol_w_mf3}
&& B^{\hat\phi}  =  \frac{1}{r\ell\left(\ell+1\right)}
	\left[\partial_{r}\left(rU_{{\ell m}}\right)\frac{1}{\sin\theta}
	\partial_{\phi}Y_{\ell m}+
	i\omega rV_{\ell m}\partial_{\theta}Y_{\ell m}
	\right]e^{-i\omega t} \ ,
\end{eqnarray}
and
\begin{eqnarray}
\label{sol_w_ef1}
&& E^{\hat r}  =  \frac{1}{r} V_{\ell m}Y_{\ell m}
	e^{-i\omega t}\ ,
\\\nonumber\\
\label{sol_w_ef2}
&& E^{\hat\theta}  = \frac{1}{r\ell\left(\ell+1\right)}
    \left[\partial_{r}\left(rV_{{\ell m}}\right)\partial_{\theta}Y_{\ell m}
    + \frac{i\omega}{\sin\theta}rU_{\ell m}\partial_{\phi}Y_{\ell m}
    \right]e^{-i\omega t} \ ,
\\\nonumber\\
\label{sol_w_ef3}
&& E^{\hat\phi}  = \frac{1}{r\ell\left(\ell+1\right)}
    \left[\partial_{r}\left(rV_{{\ell m}}\right)\frac{1}{\sin\theta}
    \partial_{\phi}Y_{\ell m}-
    i\omega rU_{\ell m}\partial_{\theta}Y_{\ell m}
    \right]e^{-i\omega t} \ .
\end{eqnarray}
It is useful to note that while no difference would be introduced in the
expressions of the radial components of the magnetic and electric fields
[{\it i.e.} equations (\ref{sol_w_mf1}) and (\ref{sol_w_ef1})], the
general-relativistic expressions for the remaining components would be
corrected by coefficients of the type $\sim N$ or $\sim 1/N$, and
effectively very small as soon as one considers regions of the spacetime
away from the stellar surface~\citep{r3}.

	In the typical configuration which we intend to consider, that is
inspired by astronomical observations, the star is endowed with a
background magnetic field that is essentially stationary on the timescale
of the stellar oscillations. The background electric fields will then
either be zero or induced by the time-variations of the background
magnetic field (such as those produced by the stellar oscillations or by
its rotation). In this case, no background electromagnetic fields will be
present in the wave-zone, and expressions
(\ref{sol_w_mf1})--(\ref{sol_w_ef3}) will then effectively refer to the
perturbations. As such they should be denoted with a symbol ``$\delta$''
but, in order to keep the expressions compact and since it is not
necessary to distinguish them from the background expressions, the symbol
``$\delta$'' will be omitted hereafter.

	Substituting the above expressions
(\ref{sol_w_mf1})--(\ref{sol_w_ef3}) into the Maxwell equations evaluated
in a flat spacetime, these can be recast into ordinary wave-like
equations for the Debye potentials $U$ (or $V$), {\it i.e.}
\begin{equation}
\label{weq}
\partial^2_{t}U - \partial^2_{r} U
	- \frac{1}{r^2\sin\theta}
	\partial_{\theta}\left(\sin\theta \partial_{\theta}U\right)-
	\frac{1}{r^2\sin^2\theta}\partial^2_{\phi}U
        =0 \ ,
\end{equation}
which would instead take the form 
\begin{equation}
\label{weq_gr}
\partial^2_{t}{\widetilde U} - N^2\partial_{r}
	\left(N^2\partial_{r}{\widetilde U}\right)
	- N^2\left[\frac{1}{r^2\sin\theta}
 	\partial_{\theta}\left(\sin\theta \partial_{\theta}
 	{\widetilde U}\right)+
 	\frac{1}{r^2\sin^2\theta}\partial^2_{\phi}{\widetilde U} \right]
         =0 \ ,
\end{equation}
on a Schwarzschild background. Note that while equations (\ref{weq}) and
(\ref{weq_gr}) are not wave equations for the potentials $U$, $V$ (or the
corresponding relativistic ones ${\widetilde U}$, ${\widetilde V}$) ,
they are so for the potentials $U_{\ell m}$, $V_{\ell m}$ (or the
corresponding relativistic ones ${\widetilde U}_{\ell m}$ or ${\widetilde
V}_{\ell m}$) .

	The main advantage of the use of a flat background spacetime for
the study of the electromagnetic fields in the wave-zone region comes
from the fact that while no analytic solution can be found for equation
(\ref{weq_gr}), an analytic solution to equation (\ref{weq}) (regular
everywhere except at $r=0$), can be expressed in terms of outgoing
spherical Hankel functions $H_{\ell}(\omega r)$ in the form
\begin{equation}
U_{\ell m}(r)=\left[\ell\left(\ell+1\right)\right]^{1/2}
	H_{\ell}(\omega r)u_{\ell m} \ , 
\hskip 3.0 cm
V_{\ell m}(r)=-\left[\ell\left(\ell+1\right)\right]^{1/2} 
	H_{\ell}(\omega r)v_{\ell m} \ ,
\end{equation}
so that the components for the magnetic fields are given by the general
expressions
\begin{eqnarray}
\label{sol_wz_mf1}
&& B^{\hat r}  =  \frac{e^{-i\omega t}
    \sqrt{\ell\left(\ell+1\right)}}{r}
    H_{\ell}(\omega r)u_{\ell m}Y_{\ell m}\ ,
\\\nonumber\\
\label{sol_wz_mf2}
&& B^{\hat\theta}  =  \frac{e^{-i\omega t}}
    {\sqrt{\ell(\ell+1)}}
    \left(DH_{\ell}(\omega r)u_{\ell m}\partial_{\theta}Y_{\ell m}-
    \omega H_{\ell}(\omega r)v_{\ell m}\frac{mY_{\ell m}}{\sin\theta}
    \right)\ ,
\\ \nonumber\\
\label{sol_wz_mf3}
&& B^{\hat\phi}  =  i\frac{e^{-i\omega t}}
    {\sqrt{\ell(\ell+1)}}
    \left(DH_{\ell}(\omega r) u_{\ell m}\frac{mY_{\ell m}}{\sin\theta}
    -\omega H_{\ell}(\omega r)v_{\ell m}\partial_{\theta}Y_{\ell m}\right)\ ,
\end{eqnarray}
while the electric field components are expressed as
\begin{eqnarray}
\label{sol_wz_ef1}
&& E^{\hat r}  =  -\frac{e^{-i\omega t}
    \sqrt{\ell\left(\ell+1\right)}}{r}
    H_{\ell}(\omega r)v_{\ell m}Y_{\ell m}\ ,
\\\nonumber\\
\label{sol_wz_ef2}
&& E^{\hat\theta}  =  -\frac{e^{-i\omega t}}
    {\sqrt{\ell(\ell+1)}}
    \left(DH_{\ell}(\omega r)v_{\ell m}\partial_{\theta}Y_{\ell m}+
    \omega H_{\ell}(\omega r)u_{\ell m}\frac{mY_{\ell m}}{\sin\theta}
    \right)\ ,
\\ \nonumber\\
\label{sol_wz_ef3}
&& E^{\hat\phi}  =  -i\frac{e^{-i\omega t}}
    {\sqrt{\ell(\ell+1)}}
    \left(DH_{\ell}(\omega r) v_{\ell m}\frac{mY_{\ell m}}{\sin\theta}
    +\omega H_{\ell}(\omega r)u_{\ell m}\partial_{\theta}Y_{\ell m}
    \right)\ .
\end{eqnarray}
Here, we have denoted the radial derivative as $DH_{\ell}(\omega r)
\equiv r^{-1}\partial_{r}\left[rH_{\ell}(\omega r)\right]$ and we recall
that the spherical Hankel functions have a simple radial fall-off in the
case of small arguments, {\it i.e.}
\begin{equation}
\label{small} 
H_{\ell}(\omega r)\approx -i(2\ell-1)!!
	(\omega r)^{-\ell-1}\ , \qquad
	DH_{\ell}(\omega r)\approx i\ell(2\ell-1)!!
	(\omega r)^{-\ell-2}\omega
	= - \ell \frac{H_{\ell}}{r}\ ,
	\qquad {\rm for} \qquad
	\omega r\approx   \omega_{_R}R\ll 1 \ ,
\end{equation}
while they exhibit a typical oscillatory behaviour (in space) in the
limit of large arguments (see, for instance, Arfken \& Weber 2001), {\it
i.e.}
\begin{equation}
\label{large}
	H_{\ell}(\omega r)\approx (-i)^{\ell+1}
	\frac{e^{i\omega r}}{\omega r}\ ,
	\qquad
	DH_{\ell}(\omega r)
	\approx (-i)^{\ell}\frac{e^{i\omega r}}{r} 
	= i \omega H_{\ell}\ , 
	\qquad {\rm for} \qquad
    \omega r\rightarrow \infty \ .
\end{equation}

	We also note that the Newtonian expressions
(\ref{sol_wz_mf1})--(\ref{sol_wz_ef3}) are not new and have indeed been
discussed by a number of authors before (see, for instance, McDermott et
al.  1984). However, new and general relativistic corrections are
introduced in equations (\ref{sol_wz_mf1})--(\ref{sol_wz_ef3}) once the
integration coefficients $u_{\ell m}$ and $v_{\ell m}$ are specified
through the matching of the electromagnetic fields
(\ref{sol_wz_mf1})--(\ref{sol_wz_ef3}) at the stellar surface with the
help of the boundary conditions (\ref{bound_e2})--(\ref{bound_e3})
\begin{eqnarray}
\label{v_coef}
&& v_{\ell m} = \frac{1}{\sqrt{\ell(\ell+1)}}
    \frac{e^{i\omega_{_R}t}}{DH_{\ell}(\omega_{_R}R)N_{_R}}
    \int d\Omega\left\{\partial_{\theta}Y^*_{\ell m}
    \left[\delta v_{_R}^{\hat r} B^{\hat \phi}_{_R}
        -\delta v_{_R}^{\hat \phi}B^{\hat r}_{_R}\right] +
        i\frac{mY^*_{\ell m}}{\sin\theta}
        \left[\delta v_{_R}^{\hat\theta} B^{\hat r}_{_R}
        -\delta v_{_R}^{\hat r}B^{\hat \theta}_{_R}\right]\right\}\ ,
\\ \nonumber\\
\label{u_coef}
&& u_{\ell m} = \frac{1}{\sqrt{\ell(\ell+1)}}
    \frac{e^{i\omega_{_R}t}}{H_{\ell}(\omega_{_R}R)N_{_R}\omega_{_R}}
    \int d\Omega\left\{i\partial_{\theta}Y^*_{\ell m}
    \left[\delta v_{_R}^{\hat r} B^{\hat \theta}_{_R}
        -\delta v_{_R}^{\hat \theta}B^{\hat r}_{_R}\right] +
        \frac{mY^*_{\ell m}}{\sin\theta}
        \left[\delta v_{_R}^{\hat\phi} B^{\hat r}_{_R}
        -\delta v_{_R}^{\hat r}B^{\hat\phi}_{_R}\right]\right\}\ , 
\end{eqnarray}
and of the relation
\begin{equation}
\label{ylms_rel}
\int d\Omega \left(
	\partial_{\theta}Y_{\ell m}\partial_{\theta}Y^*_{\ell m} - 
	\frac{m Y_{\ell m}}{\sin\theta}\frac{m Y^*_{\ell m}}{\sin\theta}
	\right) = - \ell (\ell + 1)
	\ .
\end{equation}

	Indeed, and as it will become apparent in the discussion made in
the following Section, all of the general relativistic effects are
embodied in the proper specification of the integration constants
(\ref{v_coef})--(\ref{u_coef}). In the case of the electromagnetic
radiation produced by a rotating dipole, for instance, they will be
responsible for the most important quantitative corrections of general
relativistic nature.

\section{A Useful Application: a Rotating Magnetic Dipole}
\label{rmd}

	In the following, we will discuss the application of the
expressions derived in Sections~\ref{em_nz} and ~\ref{em_wz} for a
configuration frequently discussed in astrophysics: the electromagnetic
fields and the corresponding energy losses produced when a relativistic
and massive magnetic dipole moment rotates in vacuum. Although idealized,
this simple model has been widely used to provide a first quantitative
description of the phenomenology associated with pulsars. The expressions
for the electromagnetic fields and for the energy losses through dipolar
electromagnetic radiation have been derived long ago within a Newtonian
description \citep{p68} and, to the best of our knowledge, have not been
extended before to a fully general relativistic framework. This complex
problem, however, can find a simple solution within the formalism
discussed so far if it is recast into the problem of determining the
electromagnetic fields in the near and wave-zone of a relativistic and
perfectly conducting stellar crust moving with a velocity field
\begin{equation}
\label{vel_uniform}
\delta u^\alpha =
        \frac{1}{N}\bigg(1, 0, 0, \Omega\bigg) \ ,
\end{equation}
and whose corresponding components of the three-velocity are
\begin{equation}
\label{3vel_uniform}
	\delta v^{\hat i}\equiv 
	\bigg(0, 0, \Omega r\sin\theta\bigg)\ , 
\end{equation}
where $r \leq R$ and $\Omega$ is the angular velocity of the star as
measured by a distant observer.

	Being axisymmetric, the velocity field (\ref{vel_uniform}) will
not introduce first-order perturbations in the near and wave-zones if the
magnetic dipole is aligned with the stellar rotation axis. However, if a
nonzero inclination angle $\chi$ is present between the stellar magnetic
dipole and the rotation axis, this velocity field will introduce a time
modulation of the magnetic and electric field components in the near-zone
as well as the emission of electromagnetic waves in the wave-zone. These
variations can be compared with the expressions for the electromagnetic
fields of a rotating relativistic magnetized star available in the
literature \citep{r1,r2,k3}, thus offering a useful testbed for the
formalism developed so far. Before doing that, however, it is necessary
to calculate the expressions for the static background dipolar magnetic
field using the general expressions (\ref{nz_mfg1})--(\ref{nz_mfg3}).

\subsection{Near-Zone Background Magnetic Fields of a Static Dipole}
\label{dipolarmf}

	Hereafter, we will calculate the dipolar properties of the
background magnetic field by considering the general expressions
(\ref{nz_mfg1})--(\ref{nz_mfg3}) with $\ell=1$, but will also allow for
nonzero values of the index $m$ to investigate the important departures
from axisymmetry. Furthermore, we will assume the magnetic dipole to be
varying on a timescale much longer than that set by the stellar
oscillations so that it can effectively be considered static. In this
case, the scalar functions $S_{1m}$ in (\ref{sol_P}) are time-independent
and can be easily calculated to be
\begin{equation}
\label{S1m}
S_{1m}(r) = \frac{r^2}{2M^3}\left[\ln N^2+
    \frac{2M}{r}\left(1+\frac{M}{r}\right)\right]
    s_{1m}\ ,
\end{equation}
after using the expression for the first Legendre function of the second
kind\footnote{Note that in Rezzolla et al. 2001a, the incorrect
expression was given for $Q_1$. In particular, equation (89) of that
paper shows the expression for $Q_0$; the correct expression for $Q_1$
was however used in the calculations reported there.}
\begin{equation}
\label{Q_1}
Q_1 = \frac{x}{2} \ln \left( \frac{x+1}{x-1} \right) - 1 \ .
\end{equation}
Substituting (\ref{S1m}) in equations (\ref{nz_mfg1})--(\ref{nz_mfg3})
yields the following components for the background magnetic field in the
near-zone
\begin{eqnarray}
\label{nz_mf1}
&& B^{\hat r} = \frac{1}{M^3}\left[\ln N^2+
    \frac{2M}{r}\left(1+\frac{M}{r}\right)\right]
    s_{1m} Y_{1m}
= \sqrt{\frac{3}{8\pi}}\frac{1}{M^3}\left[\ln N^2 +
    \frac{2M}{r} \left(1 +  \frac{M}{r}\right) \right]
    \left[ \sqrt{2}s_{10}   \cos\theta
    - \mathfrak{Re} \left(s_{11}
    e^{i\phi}\right) \sin\theta \right]\ ,
\nonumber \\ \\
\label{nz_mf2}
&& B^{\hat \theta} =\frac{N}{M^2r}\left[
        \frac{r}{M}\ln N^2+
    \frac{1}{N^2}+1\right]
     s_{1m}
    \partial_{\theta}  Y_{1m} 
= -\sqrt{\frac{3}{8\pi}}
    \frac{N}{M^2r}\left[\frac{r}{M}\ln N^2+
    \frac{1}{N^2}+1\right]
    \left[\sqrt{2}s_{10}    \sin\theta
    + \mathfrak{Re}\left(s_{11}e^{i\phi}\right) \cos\theta \right]\ ,
\nonumber \\ \\
\label{nz_mf3}
&& B^{\hat \phi} =\frac{N}{M^2r\sin\theta}
    \left[\frac{r}{M}\ln N^2+
    \frac{1}{N^2}+1\right] s_{1m}
    \partial_\phi  Y_{1m}
= -\sqrt{\frac{3}{8\pi}}
    \frac{N}{M^2r}\left[\frac{r}{M}\ln N^2+
    \frac{1}{N^2}+1\right]
    \left[\mathfrak{Re}\left( i s_{11}e^{i\phi}\right) \right]\ ,     
\end{eqnarray}
where we have chosen to consider $s_{1m}=0$ for $m < 0$ and where
$\mathfrak{Re}(A)$ is the real part of $A$.

	Expressions (\ref{nz_mf1})--(\ref{nz_mf3}) are the general
relativistic vacuum magnetic field components in the near-zone of a
relativistic spherical star with a dipolar magnetic field. To be fully
determined, they further need the specification of the coefficients
$s_{1m}$ which can be calculated through the use of suitable boundary
conditions across the stellar surface. As mentioned in the previous
Section, these integration constants represent the way in which the
information about the properties of the star, and their general
relativistic corrections, are impressed upon the electromagnetic fields
both in the near-zone and in the wave-zone. One way of determining these
boundary conditions could be by ensuring the continuity of the radial
component of the magnetic field across the stellar surface, but this
would require knowledge of the value of the magnetic field components at
the stellar surface and these are not necessarily easy to
measure. Alternatively, the coefficients can be computed after
multiplying both sides of (\ref{nz_mf1}) by $Y^{*}_{\ell' m'}$ and
exploiting the orthogonality of the spherical harmonics: {\it i.e.} $\int
d\Omega Y_{\ell m} Y^*_{\ell' m'} = \delta_{\ell \ell'} \delta_{m
m'}$. Doing so yields the following expressions for the unknown
coefficients
\begin{equation}
\label{s_1m}
s_{1m} = \frac{M^3 r^2 }{r^2 \ln N^2+
    2M\left(r + M\right)} \int B^{\hat r} Y^{*}_{1m} d\Omega \ ,
\end{equation}
which must hold at any radius $r$ and, in particular, in the
asymptotically flat portion of the spacetime, where $M/r \ll 1$ and the
magnetic field components assume their Newtonian expressions
\begin{equation}
\label{mf_newtonian}
\bigl(B^{\hat i}\bigr)_{_{\rm Newt.}} = 
	\frac{B_0 R^3}{2 r^3}\Bigg(
	2\left(\cos\chi\cos\theta + \sin\chi\sin\theta\cos\phi\right),~
	\cos\chi\sin\theta - \sin\chi\cos\theta\cos\phi,~
	       \sin\chi\sin\phi
	\Bigg) \ ,
\end{equation}
where, $B_0 \equiv 2 \mu/R^3$ is the (Newtonian) value of the magnetic
field at the polar axis. Exploiting this property, the integrals on the
right-hand-side of equation (\ref{s_1m}) can be computed analytically
giving
\begin{equation}
\label{slm}
s_{10}=-\frac{\sqrt{3\pi}}{4} B_0 R^3\cos\chi \ , \hskip 2.0 cm
s_{11}= \sqrt{\frac{3\pi}{2}} \frac{B_0 R^3}{2}\sin\chi\ .
\end{equation}
In the limit of flat spacetime, {\it i.e.} for $M/r\rightarrow 0$ and
$M/R\rightarrow 0$, expressions (\ref{nz_mf1})--(\ref{nz_mf3}) give
\begin{eqnarray}
\label{limit_B_1}
&& \lim_{M/r\rightarrow 0, M/R\rightarrow 0} B^{\hat r}=
    -\frac{8}{3r^3}s_{1m}Y_{1m} =
        \left(\frac{R}{r}\right)^3 B^{\hat r}_{_R} \ ,
\\ \nonumber \\
\label{limit_B_2}
&& \lim_{M/r\rightarrow 0,  M/R\rightarrow 0} B^{\hat\theta}=
    \frac{4}{3r^3}s_{1m}\partial_{\theta}Y_{1m} =
        -\frac{1}{2}\left(\frac{R}{r}\right)^3
        \partial_{\theta} B^{\hat r}_{_R} \ ,
\\ \nonumber \\
\label{limit_B_3}
&& \lim_{M/r\rightarrow 0, M/R\rightarrow 0} B^{\hat\phi}=
    \frac{4}{3r^3\sin\theta}s_{1m}\partial_{\phi}Y_{1m} =
        -\frac{1}{2}\left(\frac{R}{r}\right)^3
        \frac{1}{\sin\theta}\partial_\phi B^{\hat r}_{_R}
         \ ,
\end{eqnarray}
where $B^{\hat r}_{_R}=B_0(\cos\chi \cos\theta + \sin\chi \sin\theta
\cos\phi)$ is the radial component of the magnetic field at the stellar
surface. As expected, expressions (\ref{limit_B_1})--(\ref{limit_B_3})
coincide with the near-zone solutions of Muslimov \& Tsygan (1986) for
the dipolar magnetic field of a Newtonian magnetized star [{\it cf.} eqs. (18)
of Muslimov \& Tsygan, 1986] and that a toroidal magnetic field can be
produced by a poloidal one when this is not axisymmetric [{\it cf.} equations
(\ref{nz_mf3}) and (\ref{limit_B_3})].

\subsection{Near-zone Electromagnetic Fields from a Rotating Magnetic Dipole} 
\label{rd_nz}

	Given the background dipolar magnetic fields discussed in the
previous Section we now examine the electromagnetic fields that are
produced when the stellar dipole is inclined at an angle $\chi$ with
respect to the stellar polar axis and rotates uniformly with an angular
velocity $\Omega$ as measured by a distant observer. In this case, the
integration constants $s_{\ell m}$ for the background magnetic field are
the same as in (\ref{slm}), while the values for the integration
constants $\delta s_{\ell m}$ relative to the perturbed magnetic field
are computed to be (see Appendix \ref{der_t_delta_s} for details)
\begin{equation}
\label{dslm}
\delta s_{10} = 0 \ , 	\hskip 2.0 cm
\delta s_{11} = \sqrt{\frac{3\pi}{2}}\frac{B_0 R^3}{2}
	(e^{-i\Omega t} - 1) \sin\chi\ ,
\end{equation}
where, again, we have chosen $\delta s_{1 m}=0$ for $m <0$. Using now
(\ref{slm}) and (\ref{dslm}) in expressions
(\ref{nz_mfg1})--(\ref{nz_mfg3}), the complete magnetic field components
produced in the near-zone by the relativistic rotating magnetic dipole
are then given by the general expressions
\begin{eqnarray}
\label{mf_1}
&& B^{\hat r} = - \frac{3 R^3}{8M^3}
    \left[\ln N^2 + \frac{2M}{r}\left(1 +  \frac{M}{r}
    \right) \right]B_0\; (\cos\chi \cos\theta +
    \sin\chi \sin\theta \cos\lambda) 
    \ ,
\\\nonumber\\
\label{mf_2}
&& B^{\hat \theta} = \frac{3 R^3 N}{8 M^2 r}
    \left[\frac{r}{M}\ln N^2 +\frac{1}{N^2}+ 1
    \right]B_0\; (\cos\chi \sin\theta
    - \sin\chi \cos\theta \cos\lambda)
    \ ,
\\\nonumber\\
\label{mf_3}
&& B^{\hat \phi} = \frac{3 R^3 N}{8 M^2 r}
    \left[\frac{r}{M}\ln N^2 +\frac{1}{N^2}+ 1
    \right]B_0\; (\sin\chi \sin\lambda)
    \ ,
\end{eqnarray}
where $\lambda \equiv \phi - \Omega t$. Although derived in a different
way, expressions (\ref{mf_1})--(\ref{mf_3}) coincide with the magnetic
field components of a magnetized relativistic star in the slow-rotation
approximation [{\it cf.} equations (97)--(99) of Rezzolla et
al. 2001a]. Furthermore, in the case in which $\chi=0$, expressions
(\ref{mf_1})--(\ref{mf_3}) simply reduce to the magnetic field components
of a relativistic dipole (Ginzburg \& Ozernoy, 1964; Anderson \& Cohen,
1970).

	We next consider the expressions for the electric field using
equations (\ref{nz_ef1})--(\ref{nz_ef3}) with $\ell=2$ for a quadrupolar
electric field and exploiting the relation between the $X_{\ell m}$ and
$S_{\ell m}$ functions [{\it cf.} equation (\ref{sol_X})] which leaves as the
only nonzero component
\begin{equation}
\label{sol_X1m}
X_{1m} = \frac{r^2}{2}\left[\ln N^2+
    \frac{2M}{r}\left(1+\frac{M}{r}\right)\right]
    \delta x_{1m} (t)\ .
\end{equation}
With some straightforward algebra, we finally obtain that the components
of the electric field produced in the near-zone by the rotating magnetic
dipole assume the form
\begin{eqnarray}
\label{nz_rot_ef1}
&& \delta E^{\hat r} = \frac{1}{3}\frac{E_0}{g_{_R}} 
    \frac{1}{N_{_R}} \left[\left(3-\frac{2r}{M}\right) \ln N^2 +
    \frac{2M^2}{3r^2}+\frac{2M}{r}-4\right] \left[\cos\chi (3
    \cos^2\theta-1)+3\sin\chi\sin\theta\cos\theta\cos\lambda \right] \ ,
    \\\nonumber \\
\label{nz_rot_ef2}
&& \delta E^{\hat \theta} = - \frac{1}{2} \frac{E_0}{g_{_R}}
    \frac{N}{N_{_R}} \left[\left(1-\frac{r}{M}
    \right)\ln N^2-2-\frac{2M^2}{3r^2N^2}\right]
    \left[2\cos\chi\sin\theta\cos\theta -
    \left(\cos^2\theta-\sin^2\theta\right)
    \sin\chi\cos\lambda\right] \ ,
\\\nonumber \\
\label{nz_rot_ef3}
&& \delta E^{\hat \phi} = - \frac{1}{2}\frac{E_0}{g_{_R}}
    \frac{N}{N_{_R}} \left[\left(1-\frac{r}{M}
    \right)\ln N^2-2-\frac{2M^2}{3r^2N^2}\right]
    \sin\chi\cos\theta\sin\lambda
    \ ,
\end{eqnarray}
where we express with $g_{_R}$ the constant coefficient
\begin{eqnarray}
\label{g_R}
&& g_{_R}\equiv
    \left(1-\frac{R}{M}\right)
    \ln N^2_{_{_R}} - \frac{2 M^2}{3R^2 N^2_{_{_R}}} - 2 = 
    \left(1-\frac{R}{M}\right)\ln \left(1-\frac{2M}{R}\right)
    - \frac{2}{3}\left(\frac{M}{R}\right)^2 \frac{R}{R-2M} - 2
    \ ,
\end{eqnarray}
and where, in analogy with the corresponding Newtonian expression, we
define the electric field $E_0$ as
\begin{equation}
E_0 \equiv (f_{_R} B_0) \frac{\Omega}{N_{_R}} R 
	= (f_{_R} B_0) \Omega_{_R} R  \ .  
\end{equation}

	It is worth noting that the coefficient $f_{_R}$ introduced in
expression (\ref{f_R}) has also an important physical meaning since it
gives a simple measure of the relativistic corrections which lead to an
effective amplification of the magnetic fields in the vicinity of the
stellar surface. In particular, if we indicate with ${\widetilde B_0}$
the radial component of the magnetic field measured at the surface of the
relativistic star for an aligned nonrotating magnetic dipole, {\it i.e.}
${\widetilde B}_0 \equiv B^{\hat r}(r=R, \theta = 0, \chi=0, \Omega =
0)$, then the ratio
\begin{equation}
\label{mf_amplfctn}
\frac{{\widetilde B}_0}{B_0} = \frac{{\widetilde B}_0 R^3}{2 \mu} = 
f_{_R} \ ,
\end{equation}
represents the general relativistic {\it amplification} of the magnetic
field strength [{\it cf.} (\ref{mf_1})]. This amplification factor
$f_{_R}$ is a direct consequence of the spacetime curvature produced by
the star and is shown in the small inset of Fig.~\ref{fig1}, where it
takes values between 1.1 and 1.5 for the values of the stellar
compactness $M/R$ usually associated with compact stars. With simple
algebra and proper order accounting, it is not difficult to show that
$f_{_R}=1$ in the Newtonian limit [{\it cf.} equation (93) of
\citet{r1}], so that $E_0=\Omega R B_0$ will then represent the
well-known electric field induced by the rotation of a Newtonian
magnetized star.

	Also for the electric fields, it is possible to show that
expressions (\ref{nz_rot_ef1})--(\ref{nz_rot_ef3}) coincide with the
solutions found for the exterior electric field of a rotating misaligned
dipole in a Schwarzschild spacetime [{\it cf.} equations (124)--(126) of
Rezzolla et al. 2001a]. Furthermore, in the limiting case of a flat
spacetime, {\it i.e.}  $M/r,~ M/R\rightarrow 0$, the expressions coincide
with the near-zone solutions of Deutsch (1955). Overall, the results
presented in this Section confirm the consistency of the procedure
followed and show the flexibility of the formalism presented here.

\subsection{Wave-zone Electromagnetic Fields from a Rotating Magnetic Dipole} 
\label{rd_wz}

	The expressions for the electromagnetic fields produced in the
wave-zone by a rotating magnetic dipole were calculated long ago in
Newtonian gravity, dating back to the work of Deutsch (1955), who
presented them in terms of a complex but complete multipolar
expansion. The extension of Deutsch's results to a general relativistic
framework using the procedure discussed in Section \ref{em_wz} is rather
straightforward, but we will consider it here only for the lowest-order
multipoles. In particular, we will discuss the dipolar ({\it i.e.} those
with $\ell =1$) and quadrupolar ({\it i.e.} those with $\ell = 2$) parts
of the electric and magnetic fields in the wave-zone since these are the
ones with the slowest radial fall-off and are therefore the ones mainly
responsible for the energy losses.

	After some lengthy but straightforward algebra in which
conditions (\ref{v_coef}) and (\ref{u_coef}) are written explicitly in
terms of the magnetic field components (\ref{mf_1})--(\ref{mf_3}) and of
the velocity field (\ref{vel_uniform}), it is possible to determine the
different integration constants $u_{\ell m}$ and $v_{\ell m}$ to be used
in expressions (\ref{sol_wz_mf1})--(\ref{sol_wz_ef3}). More specifically,
when $\ell = 1$ one finds that $v_{1m}=0$ and the only nonzero
coefficient is given by $u_{11}$ and has an explicit expression

\begin{equation}
\label{u_11}
u_{11}=-i\sqrt{\frac{4\pi}{3}}  
	\Omega^2_{_R} R^3 f_{_R} B_0 \sin\chi\ . 
\end{equation}
Using this result, the dipolar parts of the electromagnetic fields
(\ref{sol_wz_mf1})--(\ref{sol_wz_ef3}) induced in the wave-zone by the
rotation of the dipole are then expressed as the real parts of the
following solutions
\begin{eqnarray}
\label{wz_ros_1}
&& B^{\hat r} =  
        -i\frac{\Omega_{_R} R^3}{N_{_R} r^2}
	f_{_R} B_0 \sin\theta
        \sin\chi {e^{i[\Omega (r-t)+\phi]}} \ ,
\\ \nonumber\\
\label{wz_ros_2}
&& B^{\hat\theta} = 
        \frac{1}{2}\frac{\Omega^2_{_R} R^3}{r}
	f_{_R} B_0 \cos\theta
	\sin\chi
	{e^{i[\Omega(r- t)+\phi]}} \ ,
\\ \nonumber\\
\label{wz_ros_3}
&& B^{\hat\phi} = 
        \frac{i}{2}\frac{\Omega^2_{_R} R^3}{r}
	f_{_R} B_0
	\sin\chi {e^{i[\Omega(r- t)+\phi]}} \ ,
\\ \nonumber\\
\label{wz_ros_4}
&& E^{\hat r} = 0
     \ ,
\\ \nonumber\\
\label{wz_ros_5}
&& E^{\hat\theta} = 
        \frac{i}{2}\frac{\Omega_{_R} R^2}{r}
	E_0
	\sin\chi {e^{i[\Omega(r- t)+\phi]}} = 
	B^{\hat \phi} \ ,
\\ \nonumber\\
\label{wz_ros_6}
&& E^{\hat\phi} = 
        -\frac{1}{2}\frac{\Omega_{_R} R^2}{r}
	E_0 \cos\theta
	\sin\chi
	{e^{i[\Omega(r- t)+\phi]}} = 
	- B^{\hat \theta} \ .
\end{eqnarray}
Since the wave-zone is located well outside the light cylinder, {\it
i.e.} at $r \gg r_{lc} \equiv 1/\Omega$, expressions
(\ref{wz_ros_1})--(\ref{wz_ros_6}) show that in this region the
electromagnetic fields behave essentially as radially outgoing waves for
which $|B^{\hat r}/B^{\hat \theta}| \sim |B^{\hat r}/B^{\hat \phi}| \sim
1/\Omega r \ll 1$.

	Proceeding in a similar manner when $\ell=2$ and requiring the
solutions to be regular at $\theta=0$, one finds that $u_{2m}=0$ and the
only nonzero coefficient is then given by $v_{21}$, whose explicit form
is
\begin{equation}
\label{u_21}
v_{21}=
	\frac{i}{3}\sqrt{\frac{\pi}{5}}  
	\Omega^4_{_R} R^5 f_{_R} B_0 \sin\chi\ . 
\end{equation}
As a result, the quadrupolar parts of the electromagnetic fields induced
in the wave-zone by the rotation of the magnetic dipole are expressed as
the real parts of the following solutions
\begin{eqnarray}
\label{wz_ros_7}
&& B^{\hat r} =  0 \ ,
\\ \nonumber\\
\label{wz_ros_8}
&& B^{\hat\theta} = 
	- \frac{1}{12}\frac{\Omega^4_{_R} R^5}{r}
	f_{_R} B_0 
  	\cos\theta\sin\chi {e^{i[\Omega(r-t) + \phi]}}
	\ ,
\\ \nonumber\\
\label{wz_ros_9}
&& B^{\hat\phi} =  
	\frac{i}{12}\frac{\Omega^4_{_R} R^5}{r}
	f_{_R} B_0 
	\left(\sin^2\theta-\cos^2\theta\right)
    	\sin\chi {e^{i[\Omega(r-t) + \phi]}}
	\ ,
\\ \nonumber\\
\label{wz_ros_10}
&& E^{\hat r} = 
	-\frac{1}{2}\frac{\Omega^2_{_R} R^4}{N_{_R} r^2}
	E_0 
	\sin\theta\cos\theta\sin\chi {e^{i[\Omega(r-t) + \phi]}}
     	\ ,
\\ \nonumber\\
\label{wz_ros_11}
&& E^{\hat\theta} = 
	\frac{i}{12}\frac{\Omega^3_{_R} R^4}{r}
	E_0 
	\left(\sin^2\theta-\cos^2\theta\right)
    	\sin\chi {e^{i[\Omega(r-t) + \phi]}}
	= B^{\hat \phi} \ , 
\\ \nonumber\\
\label{wz_ros_12}
&& E^{\hat\phi} = 
	\frac{1}{12}\frac{\Omega^3_{_R} R^4}{r}
	E_0 
  	\cos\theta\sin\chi {e^{i[\Omega(r-t) + \phi]}}
	= - B^{\hat \theta} \ . 
\end{eqnarray}
Also in this case the electromagnetic fields behave essentially as
radially outgoing waves for which $|E^{\hat r}/E^{\hat \theta}| \sim
|E^{\hat r}/E^{\hat \phi}| \sim 1/\Omega r \ll 1$.

	A few aspects of expressions (\ref{wz_ros_1})--(\ref{wz_ros_12})
are worth noting. The first and most obvious one is that all of these
components are identically zero when the magnetic dipole is aligned with
the rotation axis ({\it i.e.} $\chi=0$) underlining that an axisymmetric
configuration cannot emit electromagnetic waves. The second one is that
all components have a phase term $\propto {\rm exp}\{i[\Omega(r-t) +
\phi]\}$ expressing the wave nature of the solutions, where the term
$\propto {\rm exp}(i\Omega r)$ is inherited from the spherical Hankel
functions once $\omega = \Omega$. In the limit of flat spacetime, these
solutions coincide with those in equations (87) and (88) of Michel \& Li
(1999) for the dipolar and quadrupolar fields, respectively. Finally,
because expressions (\ref{wz_ros_1})--(\ref{wz_ros_6}) and
(\ref{wz_ros_7})--(\ref{wz_ros_12}) represent the dominant
electromagnetic fields in the wave-zone, they provide the largest
contributions to the electromagnetic losses which will be computed in the
following Section.

\subsection{Electromagnetic Radiation Losses from a Rotating Magnetic Dipole}
\label{oscilltn_dampng}

Using the results of the previous Section, we can estimate the luminosity
carried away by the dipolar electromagnetic radiation $L_{em}$ in terms
of the integral of the radial component of the Poynting vector ${\vec P}$
\begin{equation}
\label{power}
L_{em} \equiv \int_{\partial \Sigma} P^{\hat r}dS = 
	\frac{1}{4\pi}\int_{\partial \Sigma} 
	\left({\vec E}\times {\vec B}\right)^{\hat r} dS\ ,   
\end{equation}
where the integral is made over the 2-sphere $\partial\Sigma$ of radius
$r\gg 1/\Omega > R$ and surface element $dS$.

	Substituting in (\ref{power}) the expressions for the electric
and magnetic fields in the wave-zone
(\ref{sol_wz_mf2})--(\ref{sol_wz_mf3}) and
(\ref{sol_wz_ef2})--(\ref{sol_wz_ef3}), it is not difficult to find that
the radial component of the Poynting vector
\begin{equation}
P^{\hat r}=\frac{1}{4\pi} \left|
    E^{\hat\phi}B^{\hat\theta}-E^{\hat\theta}B^{\hat\phi}
    \right|=\frac{1}{4\pi\ell(\ell+1)} \left|\omega DH_{\ell}(\omega
    r)H_{\ell}(\omega r)
    \left[ \left|u_{\ell m}\right|^2+\left|v_{\ell m}\right|^2\right]
    \left(\partial_{\theta}Y_{\ell m} \partial_{\theta}Y^*_{\ell
    m}+\frac{mY_{\ell m}} {\sin\theta}\frac{mY^*_{\ell
    m}}{\sin\theta}\right)\right|\ ,
\end{equation}
%
and is basically determined by the square of the integration constants
$u_{\ell m}$ and $v_{\ell m}$ [{\it cf.} equations (\ref{v_coef}),
(\ref{u_coef})]. In the case of a purely dipolar radiation, $u_{11}$ is
the only nonzero coefficient [{\it cf.} equations (\ref{u_11})] and the power
(\ref{power}) radiated in the form of dipolar electromagnetic radiation
is then
\begin{equation}
\label{dipole_energy_loss}
L_{em} = \frac{1}{8\pi}
	\left(\left|u_{\ell m}\right|^2+\left|v_{\ell m}\right|^2\right) =
	\frac{|u_{11}|^2}{8\pi} = \frac{\Omega^4_{_R} R^6 {\widetilde
	B}^2_0}{6 c^3}\sin^2\chi
	\ .
\end{equation}
%
When compared with the equivalent Newtonian expression for the rate of
electromagnetic energy loss through dipolar radiation \citep{ll87,p68},
\begin{equation}
\label{dipole_energy_loss_newt}
(L_{em})_{\rm Newt.} = 
	\frac{\Omega^4 R^6 B_0^2}{6 c^3}\sin^2\chi 
	\ , 
\end{equation}
it is easy to realize that the general relativistic corrections emerging
in expression (\ref{dipole_energy_loss}) are due partly to the magnetic
field amplification at the stellar surface [{\it i.e.} ${\widetilde B}_0
= f_{_R} B_0$; {\it cf.} equation (\ref{mf_amplfctn})] and partly to the
increase in the effective rotational angular velocity produced by the
gravitational redshift [{\it i.e.} $\Omega = \Omega_{_R} N_{_R}$;
{\it cf.} equation (\ref{grs})]. Overall, therefore, the presence of a curved
spacetime has the effect of increasing the rate of energy loss through
dipolar electromagnetic radiation by an amount which can be easily
estimated to be
\begin{equation}
\label{dipole_energy_loss_cf}
\frac{L_{em}}{(L_{em})_{_{\rm Newt.}}}= 
    	\left(\frac{f_{_R}}{N^2_{_R}}\right)^2\ , 
\end{equation}
and whose dependence is shown in Fig.~\ref{fig1} with a solid
line. Although expression (\ref{dipole_energy_loss}) has been obtained
neglecting the curvature effects for the electromagnetic fields in the
wave-zone region ({\it cf.} Section~\ref{em_wz}), it is straightforward to
realize that already in this simplified setup, the Newtonian expression
(\ref{dipole_energy_loss_newt}) underestimates the electromagnetic
radiation losses of a factor which is between 2 and 6 in the typical
range for the compactness of relativistic stars. Furthermore, when
referred to a typical neutron star with magnetic field $B_0 = 10^{12}$ G,
$R = 12$ km, and angular velocity $\Omega=500$ rad s$^{-1}$, expression
(\ref{dipole_energy_loss}) gives an energy-loss rate
\begin{equation}
\label{dipole_energy_loss_estim}
L_{em}\approx 1.1\times 10^{39} 
	\left(\frac{f^2_{_R}}{N^4_{_R}}\right)
	\left(\frac{R}{1.2 \times 10^6\; {\rm cm}}\right)^{6}
	\left(\frac{\Omega}{500\; {\rm rad}^{-1}}\right)^{4}
	\left(\frac{B_0}{10^{12}\; {\rm G}}\right)^2
	\sin^2\chi\hskip 0.5 cm {\rm ergs~s^{-1}}\ , 
\end{equation}
where $f^2_{_R}/N^4_{_R} \simeq 4$ for $M = 1.4 M_{\odot}$.

	The expression for the energy loss (\ref{dipole_energy_loss}) can
also be used to determine the spin-evolution of a pulsar that is
converting its rotational energy into electromagnetic
radiation. Following the simple arguments proposed more than thirty years
ago \citep{p68,go69}, it is possible to relate the electromagnetic energy
loss $L_{em}$ directly to the loss of rotational kinetic energy $E_{\rm
rot}$ defined as
\begin{equation}
\label{rot_ent}
E_{\rm rot} \equiv \frac{1}{2} 
	\int d^3{\bf x} \sqrt{\gamma} e^{-\Phi(r)}
	\rho (\delta v^{\hat \phi})^2
	 \ ,  
\end{equation}
where $\rho$ is the stellar energy density. Introducing now the general
relativistic moment of the inertia of the star
\begin{equation}
\label{mom_inrt}
{\widetilde I} \equiv \int d^3{\bf x} \sqrt{\gamma} e^{-\Phi(r)} 
	\rho r^2 \sin^2\theta \ ,
\end{equation}
whose Newtonian limit gives the well-known expression for a spherical
distribution of matter $I \equiv ({\widetilde I})_{\rm Newt.} =
\frac{2}{5} M R^2$, the energy budget is then readily written as
\begin{equation}
\label{en_budgt}
{\dot E}_{\rm rot} \equiv 
	\frac{d}{dt}\left(\frac{1}{2} {\widetilde I} \Omega^2\right)
	= - L_{em} \ .  
\end{equation}
Of course, in enforcing the balance (\ref{en_budgt}) we are implicitly
assuming all the other losses of energy ({\it e.g.} those to
gravitational waves) to be negligible. This can be a reasonable
approximation except during the initial stages of the pulsar's life,
during which the energy losses due to emission of gravitational radiation
will dominate because of the steeper dependence on the angular velocity
({\it i.e.} ${\dot E}_{_{\rm GW}}\propto \Omega^6$).

\begin{figure}
\begin{center}
\includegraphics[width=10.cm,angle=0]{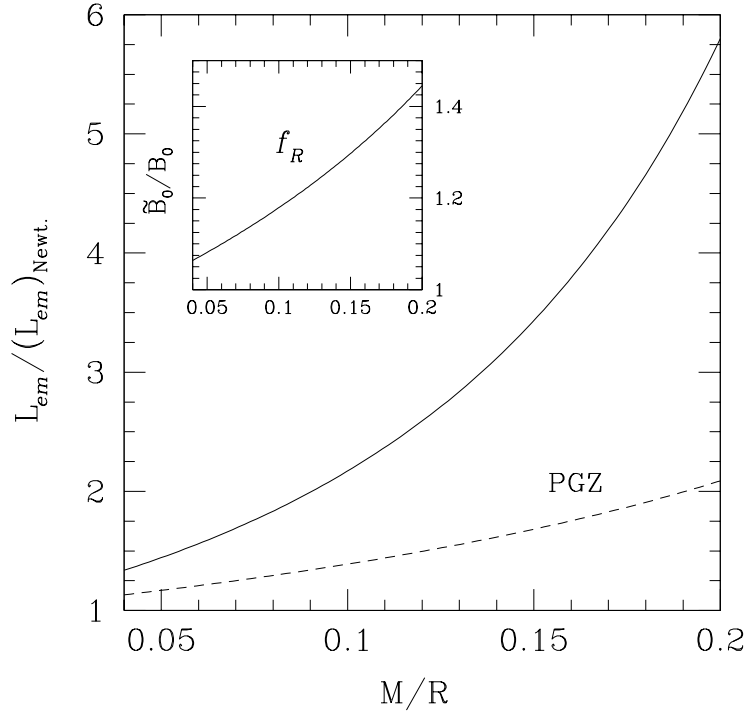}
\caption{General relativistic amplification of the energy loss through
dipolar electromagnetic radiation as a function of stellar
compactness. The solid line refers to the result derived here
[{\it cf.} eqs. (\ref{dipole_energy_loss}) and
(\ref{dipole_energy_loss_newt})], while the dashed line refers to the
phenomenological estimate made by Page et al. (2000) (PGZ). Shown in the
inset is the magnetic field amplification produced by the spacetime
curvature for different values of stellar compactness
[{\it cf.} eq. (\ref{mf_amplfctn})].}
\label{fig1}
\end{center}
\end{figure}

	Expression (\ref{en_budgt}) can also be written in a more useful
form in terms of the pulsar's most important observables: the period $P$
and its time derivative ${\dot P}\equiv dP/dt$. In this case, in fact,
using expression (\ref{dipole_energy_loss}) and (\ref{en_budgt}), it is
not difficult to show that
\begin{equation}
\label{ppdot}
P {\dot P} = \left(\frac{2 \pi^2}{3 c^3}\right) 
	\frac{1}{N^4_{_R}}
	\frac{R^6 {\widetilde B}^2_0}{\widetilde I}
	= \left(\frac{f^2_{_R}}{N^4_{_R}}\frac{I}{{\widetilde I}}\right)
	(P {\dot P})_{\rm Newt.} 
	\ ,
\end{equation}
where the Newtonian expression is given by \citep{go69}
\begin{equation}
\label{ppdot_N}
(P {\dot P})_{\rm Newt.} \equiv 
	\left(\frac{2 \pi^2}{3 c^3}\right) 
	\frac{R^6 B^2_0}{I}
	\ .
\end{equation}
Also in this case it is not difficult to realize that general
relativistic corrections will be introduced through the amplification of
the magnetic field and of the stellar angular velocity, as well as of the
stellar moment of inertia ({\it i.e.} ${I}/{{\widetilde I}}$).

	Expressions (\ref{dipole_energy_loss}) and (\ref{ppdot}) could be
used to investigate the rotational evolution of magnetized neutron stars
with predominant dipolar magnetic field anchored in the crust, either
when isolated \citep{pgz00} or when present in a binary system
\citep{letal04}. Indeed, a detailed and accurate investigation of this
type has already been performed by Page et al. (2000), who have paid
special attention to the general relativistic corrections that need to be
included for a correct modelling of the thermal evolution but also of the
magnetic and rotational evolution. It should remarked, however, that in
their treatment Page et al. (2000) have adopted an expression for the
loss of rotational kinetic energy which is similar to expression
(\ref{ppdot}) in that it accounts for the magnetic field amplification
due to the curved background spacetime, but that does not include the
corrections due to the gravitational redshift. As a result, the general
relativistic electromagnetic luminosity estimated by Page et al. (2000)
is smaller than the one computed here and it is shown with a dashed line
in Fig.~\ref{fig1} for comparison (labelled as PGZ). As the authors
themselves discuss, their choice was based on simple phenomenological
considerations and was made in the absence of a more systematic treatment
of the electromagnetic fields produced by a rotating and magnetized
relativistic star. We hope that the results reported here will be of help
in all of those investigations that consider the thermal, magnetic and
rotational evolution of magnetized neutron stars.

\section{Conclusions}
\label{conclusions}

	There is undoubted observational evidence that strong
electromagnetic fields are often present in the vicinity of relativistic
stars. At the same time, it is natural to expect that these stars will be
subject to perturbations of various types which will induce oscillations
responsible for the emission of both electromagnetic and gravitational
waves. Clearly, the possibility of detecting simultaneously both types of
signal will provide important information about the mass and radius of
these objects and hence give the ability to investigate the properties of
matter at nuclear densities. Building on the general relativistic
treatment of electrodynamics of magnetized neutron stars developed by
Rezzolla et al. (2001a,b), we have studied here the electromagnetic
fields generated when the perfectly conducting crust of a magnetized
relativistic star possesses a nonzero velocity field.  The star has been
modelled as a relativistic polytrope with infinite conductivity and in
vacuum. The relativistic corrections induced by a rotation in the
spacetime as well as those produced by the electromagnetic energy
densities have been neglected. We expect both of these approximations to
be satisfactory for the large majority of rapidly rotating and magnetized
neutron stars.

	As in our previous investigations \citep{r1,r2}, we have paid
special care to providing expressions for the electromagnetic fields that
besides being analytical, are written in a form which is as simple as
possible, that highlights the relation between the general relativistic
and Newtonian expressions, and that can therefore be used also in
astrophysical applications. This first paper has been mostly devoted to
the development of the mathematical formalism which distinguishes the
solution near the stellar surface (where the electromagnetic fields are
quasi-stationary), from the one at very large distances (where the
electromagnetic fields approach a wave-like behaviour). In both of these
regions and with suitable approximations, in fact, exact solutions to the
general relativistic Maxwell equations can be found and have shown to
contain two main corrections when compared with the corresponding
Newtonian expressions. The first one is related to the amplification of
the electromagnetic fields on the stellar surface caused by the
background spacetime curvature, while the second one is due to the
gravitational redshift which will affect the electromagnetic waves as
these propagate in the curved spacetime. It should be noted, in fact,
that while the wave-zone solutions to the Maxwell equations formally
coincide with the ones derived in a flat spacetime, integration constants
defined at the surface of the star, imprint important general
relativistic corrections which, in turn, convey information about the
mass and radius of the oscillating star.

	The general expressions for the electromagnetic fields presented
here do not refer to any specific magnetic field topology or velocity
field and are, in this sense, completely general. However, to validate
the expressions derived and to consider an astrophysically important
testbed, we have also examined the form which these electromagnetic
fields assume when the background magnetic fields are of dipolar type and
the crustal velocity field is that produced by the uniform rotation of
the dipole. This configuration is often used as the simplest model for
the electromagnetic emission observed from pulsars and has been
extensively investigated in the past. While analytic expressions are
available within a Newtonian framework both in the near-zone and in the
wave-zone, the general relativistic expressions are limited to the
near-zone. A direct comparison with these expressions and the perfect
agreement found have shown that the formalism developed here is flexible
and robust.  Furthermore, the use of analytic expressions for the
components of the electromagnetic fields in the wave-zone has allowed to
calculate, for the first time, the general relativistic expression for
the electromagnetic energy loss through dipolar radiation. A simple
comparison with the corresponding Newtonian expression shows that the
latter under-estimates this loss by a factor of 2--6, depending on the
stellar compactness.

	Work is now in progress to apply the formalism presented here to
those spheroidal and toroidal velocity fields most frequently encountered
in the study of stellar oscillations. The results will be presented in a
forthcoming companion paper \citep{r3}.

\section*{Acknowledgments}

	It is a pleasure to thank Ulrich Geppert, John Miller and Valeria
	Ferrari  for useful  discussions  and for  carefully reading  the
	manuscript. Financial support for this research has been provided
	by  the MIUR  and by  the EU  Network Program  (Research Training
	Network  Contract HPRN-CT-2000-00137). BA  also thanks  SISSA and
	INFN, Sezione di Trieste, for hospitality and financial support.

\appendix

\section[]{Derivatives of Legendre functions}
\label{Legendre}

	In this Appendix we briefly sketch the lenghty but simple algebra
needed for the calculation of the functions $Q_{\ell m}$ and their radial
derivatives, and which are used in the calculation of the electric field
in the near-zone. Using the recurrence formulae (see, for instance,
Jeffrey 1995)
\begin{eqnarray}
&& \left(x^2-1\right)\frac{dQ_{\ell}}{dx} =
    \ell\left(xQ_{\ell} - Q_{\ell-1}\right)\ ,
\\ \nonumber \\
&& \left(\ell+1\right)Q_{\ell+1}-\left(2\ell + 1\right)x
    Q_{\ell}+\ell Q_{\ell-1} = 0
\end{eqnarray}
where, we recall, $x\equiv 1 - r/M$, it is possible to show that the
radial second order derivative of Legendre functions of the second kind
takes the form
\begin{eqnarray}
\frac{d}{dx}\left[\left(1+x\right)\frac{d Q_{\ell}}{dx}\right]
    &=& \frac{\ell}{\left(x-1\right)^2}
    \left\{\left[\left(x-1\right)\ell - 1\right]Q_{\ell}
    + Q_{\ell -1}\right\}\ .
\end{eqnarray}
As a result, the expressions contained in the electric field components
(\ref{nz_ef1})--(\ref{nz_ef3}) can be written explicitly as
\begin{eqnarray}
\frac{d}{dr}\left\{r^2\frac{d}{dx}\left[\left(1+x\right)
    \frac{d Q_{\ell}}{dx}\right]\right\} &=&
         -M\frac{d}{dx}\left\{\left(x-1\right)^2\frac{d}{dx}
         \left[\left(1+x\right)
    \frac{d Q_{\ell}}{dx}\right]\right\} =
    -\frac{M\ell^2\left(\ell+1\right)}{\left(x+1\right)}
    \left(xQ_{\ell} - Q_{\ell -1}\right)\ .
\end{eqnarray}

\section[]{Calculation of the integration constants
$\partial_{t} \delta s_{\ell~m}$}
\label{der_t_delta_s}

	Inserting into (\ref{delta_B}) the value of the magnetic field
(\ref{sol_mf1})-(\ref{sol_mf3}) at the surface of the star
\begin{equation}
\label{mf_R}
B^{\hat r}_R = -\frac{1}{R^2}(\nabla^2_{_{\Omega}} S )\big\vert_{r=R}
	\ , 
	\quad B^{\hat\theta} =
	\frac{N_{_R}}{R}
	(\partial_{\theta}\partial_{r}S)\big\vert_{r=R}\ , 
	\quad B^{\hat\phi} =
	\frac{N_{_R}}{R\sin\theta}
	(\partial_{\phi}\partial_{r}S)\big \vert_{r=R} \ ,
\end{equation}
and making use of expression (\ref{sol_s_mf1}), it is possible to derive
that 
\begin{eqnarray}
\label{b2}
&& (\partial_t \delta S_{\ell m}Y_{\ell m})\big \vert_{r=R}=
    \frac{1}{\ell(\ell+1)}\Bigg\{
    (\nabla^2_{_{\Omega}} S) \frac{1}{R\sin\theta}\left[
    \partial_\theta \left(\sin\theta \delta v^{\hat\theta}\right) +
    \partial_\phi \delta v^{\hat \phi}\right]
\nonumber\\\nonumber\\
    &&+\left[N_{_R}\partial_r
    \left(\nabla^2_{_{\Omega}} S\right)\delta v^{\hat r}
    +\frac{1}{R}\partial_\theta
    \left(\nabla^2_{_{\Omega}} S\right)\delta v^{\hat\theta}
    + \frac{1}{R\sin\theta}\partial_\phi
    \left(\nabla^2_{_{\Omega}} S\right)\delta v^{\hat\phi}\right]+
    N_{_R}\left[\partial_\theta\partial_r 
    S\partial_\theta \delta v^{\hat r}+
    \frac{1}{\sin^2\theta}\partial_\phi\partial_r S
    \partial_\phi \delta v^{\hat r}\right]\Bigg\} {\Bigg\vert_{r=R}}\ ,
\end{eqnarray}
where $\nabla^2_{_{\Omega}}$ is the angular part of the Laplacian, {\it i.e.}
\begin{equation}
\nabla^2_{_{\Omega}}\equiv\frac{1}{\sin\theta}\partial_\theta
\left(\sin\theta\partial_\theta\right)+
\frac{1}{\sin^2\theta}\partial^2_{\phi}\ .
\end{equation}
	
	It is interesting to note that in the limit of flat spacetime our
expression (\ref{b2}) coincides with equation (B2) of Timokhin et
al. 2000, which was instead derived after requiring the continuity of the
tangential electric field at the stellar surface.  Multiplying this
equation by $Y^*_{\ell m}$, integrating it over the solid angle and using
equation (\ref{sol_P}), we can finally obtain an expression for the
coefficients $\partial_t (\delta S_{\ell m})$ in the expansion of $\delta
S$ in spherical harmonics
\begin{eqnarray}
\label{coef_perturb}
&& \partial_t (\delta S_{\ell m}) \vert_{r=R}=
	\frac{\ell}{M}\left\{-\left(1+\frac{R}{M}\ell\right)
	Q_{\ell} + Q_{\ell -1}\right\}\bigg\vert_{r=R}
        \partial_t \delta s_{\ell m}(t)=
\nonumber\\\nonumber\\
&& \hskip 2.5 cm
	\frac{1}{\ell(\ell+1)}\int d\Omega Y^*_{\ell m}\Bigg\{
	(\nabla^2_{_{\Omega}} S) \frac{1}{R\sin\theta}\left[
	\partial_\theta \left(\sin\theta \delta v^{\hat\theta}\right) +
	\partial_\phi \delta v^{\hat \phi}\right]
	+N_{_R}\left[(\partial_\theta\partial_r S)\partial_\theta 
	\delta v^{\hat r}+
	\frac{1}{\sin^2\theta}(\partial_\phi\partial_r S)
	\partial_\phi \delta v^{\hat r}\right]
\nonumber\\\nonumber\\
&& \hskip 2.5 cm
	+\left[N_{_R}\partial_r
	\left(\nabla^2_{_{\Omega}} S\right)\delta v^{\hat r}
	+\frac{1}{R}\partial_\theta\left(\nabla^2_{_{\Omega}} S\right)
	\delta v^{\hat\theta}
	+ \frac{1}{R\sin\theta}\partial_\phi
	\left(\nabla^2_{_{\Omega}} S\right)\delta v^{\hat\phi}\right]
	\Bigg\}\Bigg\vert_{r=R}\ .
\end{eqnarray}

\section[]{A different approach to the gravitational redshift}
\label{wave_equation}

	We here provide a somewhat different derivation of the well-known
result that photons in a gravitational potential undergo a gravitational
redshift. For doing this we will exploit the existence in this spacetime
of a timelike Killing vector $\xi^\alpha$ such that
\begin{equation}
\label{ke}
\xi_{\alpha ;\beta}+\xi_{\beta;\alpha}=0\ .
\end{equation}

	Electromagnetic waves propagate along geodesics and the
associated null wave-vector $k^\alpha$ will be tangent to these
trajectories and parallel transported along them, {\it i.e.}
\begin{equation}
\label{ge}
k^{\alpha}_{~~;\beta}k^\beta=0 \ .  
\end{equation}
We now introduce two important frequencies: the first one is the
frequency measured by an observer with four-velocity $u^{\alpha}$ and
defined as
\begin{equation}
\label{of}
\omega \equiv -k^\alpha u_\alpha \ ,  
\end{equation}
while the second one is the frequency associated with the timelike
Killing vector $\xi^{\alpha}$ and defined as
\begin{equation}
\label{kf}
\omega_\xi \equiv -k^\alpha\xi_{\alpha} \ .
\end{equation}

	The two frequencies have marked differences: while (\ref{of})
depends on the observer chosen and is therefore a function of position,
(\ref{kf}) is a conserved quantity that remains constant along the
trajectory followed by the electromagnetic wave. This is easily verified
after using equations (\ref{ke}) and (\ref{ge}) which give
\begin{equation}
\left(k^\alpha\xi_{\alpha}\right)_{,\beta}k^\beta=
	\left(k^\alpha\xi_{\alpha}\right)_{;\beta}k^\beta=
	k^\alpha_{~~;\beta}k^\beta\xi_{\alpha} +
	\xi_{ \alpha;\beta}k^\alpha k^\beta =0\ .
\end{equation}

	We can now exploit this property to measure how the frequency
changes with position and is therefore redshifted in the spacetime
(\ref{schw}). Assume the Killing vector to have components
\begin{equation}
\label{killing}
\xi^{\alpha}\equiv \bigg(1,0,0,0\bigg) \ ; \hskip 2.0cm
        \xi_{\alpha}\equiv N^2 \bigg(- 1,0,0,0 \bigg) \ ,
\end{equation}
so that $\omega_{\xi}=-k_0=$const. The frequency of an electromagnetic
wave emitted at the surface of the star $r=R$ and measured by an observer
with four-velocity $u^{\alpha}$ parallel to $\xi^{\alpha}$ ({\it i.e.} a
static observer) will be
\begin{equation}
\omega_{_R}=-(k_\alpha u^\alpha)\vert_{r=R}= N_{_R}^{-1}\omega_\xi \ ,
\end{equation}
so that at a generic radial position $r$
\begin{equation}
\label{rs}
N(r)\omega(r)= {\rm const.} = N_{_R}\omega_{_R} \ .
\end{equation}
This expression coincides with equation (\ref{grs}) in the text.

\section[]{Maxwell Equations in a Spacetime with a Rotation}
\label{Lie_derivative}

If the spacetime admits a rotational Killing three-vector $\vec \beta$,
the full set of Maxwell equations can then be written as (Thorne et
al. 1986, Rezzolla et al. 2001a, b) 
\begin{eqnarray}
\label{me_membrane_1}
&& {\vec \nabla} \cdot {\vec B} = 0\ ,
\\ \nonumber \\
&& {\vec \nabla} \cdot {\vec E} = 4\pi \rho_e\ , 
\\ \nonumber \\
&& (\partial_t - {\cal L}_{\vec
\beta}){\vec B} = - {\vec \nabla} \times (N {\vec E})\ ,
\\ \nonumber \\
\label{me_membrane_3}
&& (\partial_t - {\cal L}_{\vec \beta}){\vec E}  
	= - {\vec \nabla} \times (N {\vec B}) - 4\pi {\vec J} \ , 
\end{eqnarray}
where, again, ${\vec{\nabla}}$ represents the covariant derivative with
respect to spatial part of the metric, ${\cal L}_{\vec \beta}$ is the Lie
derivative along $\vec \beta$, and $N$ is the lapse. The system of
equations (\ref{me_membrane_1})--(\ref{me_membrane_3}) is completed with
Ohm's law, which takes the general form
\begin{equation}
{\vec j} = \sigma  \left[{\vec E} +
	\left( {\vec v} + {\vec \beta} \right) \times {\vec B}\right] \ .
\end{equation}

	The Lie derivative of a three-vector ${\vec A}$ along the vector
field ${\vec \beta}$ is not particularly difficult to calculate. However,
the form of its components in a spacetime with line element (\ref{schw})
and a frame (\ref{tetrad_0})--(\ref{tetrad_3}) is not easy to find in the
literature. For this reason we give them here as a useful reference
\begin{eqnarray}
\label{lie_1}
&&\left({\cal L}_\beta {\vec{A}}\right)^{\hat r} = 
	e^{-\Lambda}\beta^{\hat r}\partial_r A^{\hat r}
	+\frac{\beta^{\hat\theta}}{r}
	\partial_\theta A^{\hat r} 
	+\frac{\beta^{\hat\phi}}{r\sin\theta}
	\partial_\phi A^{\hat r}-  
	e^{-\Lambda}A^{\hat r}\partial_r \beta^{\hat r}
	-\frac{A^{\hat\theta}}{r}
	\partial_\theta \beta^{\hat r} 
	-\frac{A^{\hat\phi}}{r\sin\theta}
	\partial_\phi \beta^{\hat r} \ ,\\ 
\nonumber\\
\label{lie_2}
&&\left({\cal L}_\beta {\vec{A}}\right)^{\hat\theta} = 
	e^{-\Lambda}\beta^{\hat r}\partial_r A^{\hat\theta}
	+\frac{\beta^{\hat\theta}}{r}
	\partial_\theta A^{\hat\theta} 
	+\frac{\beta^{\hat\phi}}{r\sin\theta}
	\partial_\phi A^{\hat\theta}-  
	e^{-\Lambda}A^{\hat r}\partial_r \beta^{\hat\theta}
	-\frac{A^{\hat\theta}}{r}
	\partial_\theta \beta^{\hat\theta} 
	-\frac{A^{\hat\phi}}{r\sin\theta}
	\partial_\phi \beta^{\hat\theta}
	+e^{-\Lambda}\frac{A^{\hat r}\beta^{\hat\theta}- 
	A^{\hat\theta}\beta^{\hat r}}{r} \ ,\\ 
\nonumber\\
\label{lie_3}
&&\left({\cal L}_\beta {\vec{A}}\right)^{\hat\phi} = 
	e^{-\Lambda}\beta^{\hat r}\partial_r A^{\hat\phi}
	+\frac{\beta^{\hat\theta}}{r}
	\partial_\theta A^{\hat\phi} 
	+\frac{\beta^{\hat\phi}}{r\sin\theta}
	\partial_\phi A^{\hat\phi}-  
	e^{-\Lambda}A^{\hat r}\partial_r \beta^{\hat\phi}
	-\frac{A^{\hat\theta}}{r}
	\partial_\theta \beta^{\hat\phi}  
	-\frac{A^{\hat\phi}}{r\sin\theta}
	\partial_\phi \beta^{\hat\phi}
	+e^{-\Lambda}\frac{A^{\hat r}\beta^{\hat\phi}- 
	A^{\hat\phi}\beta^{\hat r}}{r} 
\nonumber\\
&& \hskip 1.5 truecm
	+\frac{A^{\hat\theta}\beta^{\hat\phi}- 
	A^{\hat\phi}\beta^{\hat\theta}}{r}\cot\theta 
	\ . 
\end{eqnarray} 
If the Lie derivative is now taken along the rotational Killing vector
\begin{equation}
\vec \beta = \frac{1}{2}\frac{g_{t\phi}}{r\sin\theta} 
	\displaystyle\boldsymbol{e}_{\phi}
	= -\omega r\sin\theta\ \displaystyle\boldsymbol{e}_{\phi} \ , 
\end{equation}
as for the spacetime of a rapidly rotating relativistic star,
expressions~(\ref{lie_1})-(\ref{lie_3}) reduce to
\begin{eqnarray}
\label{lie_1m}
&&\left({\cal L}_\beta {\vec{A}}\right)^{\hat r} = 
	- \omega \partial_\phi A^{\hat r}\ ,\\ 
\nonumber\\
\label{lie_2m}
&&\left({\cal L}_\beta {\vec{A}}\right)^{\hat\theta} = 
	- \omega \partial_\phi A^{\hat\theta}\ ,\\ 
\nonumber\\
\label{lie_3m}
&&\left({\cal L}_\beta {\vec{A}}\right)^{\hat\phi} = 
	A^{\hat r} e^{-\Lambda}\sin\theta\partial_r\omega - 
	\omega \partial_\phi A^{\hat\phi}\ . 
\end{eqnarray}

\label{lastpage}

\end{document}